\documentclass[preprint, 11pt]{article} 

\input{epsf}
\usepackage[utf8]{inputenc}
\usepackage{amsmath}
\usepackage{amssymb}
\input{epsf}
\usepackage[margin=1in]{geometry}
\usepackage[raggedright]{titlesec}
\usepackage[hang,flushmargin]{footmisc} 
\usepackage{graphicx}
\usepackage{framed}
\usepackage{color}
\usepackage{yfonts}
\usepackage[dvipsnames]{xcolor}
\usepackage{pifont}
\usepackage{amsmath,amssymb}
\usepackage{multirow}
\usepackage{amsfonts}
\usepackage{subfigure}
\usepackage{caption}
\usepackage{threeparttable}
\usepackage{esvect}
\usepackage{pgf}
\usepackage{tikz}
\usepackage{tikz-cd}
\usepackage{fancyvrb}
\usetikzlibrary{positioning}
\usepackage{bbm}
\usepackage{colortbl}
\usepackage{pdflscape}
\usepackage{booktabs}

\usepackage{amsmath}
\usepackage{mathdots}
\usepackage{yhmath}
\usepackage{cancel}
\usepackage{siunitx}
\usepackage{array}
\usepackage{multirow}
\usepackage{gensymb}
\usepackage{tabularx}
\usepackage{booktabs}
\usetikzlibrary{fadings}
\usetikzlibrary{patterns}
\usetikzlibrary{shadows.blur}
\usetikzlibrary{shapes}

\usepackage{hyperref}
\hypersetup{
    colorlinks=true,
    linkcolor=NavyBlue,
    filecolor=magenta,      
    urlcolor=cyan,
    citecolor=black
}
 
\usepackage{rotating}
\usepackage{ntheorem}
\makeatletter
\renewtheoremstyle{plain} 
{\item{\theorem@headerfont ##1\ ##2\theorem@separator}~}
{\item{\theorem@headerfont ##1\ ##2\ (##3)\theorem@separator}~}
\makeatother

\theoremheaderfont{\normalfont\bfseries}

\newtheorem{theorem}{Theorem}[section]
\newtheorem{definition}{Definition}[section]

\newtheorem{corollary}{Corollary}[section]
\newtheorem{example}{Example}[section]
\newtheorem{assumption}{Assumption}

\newcommand{\E}{\mathbb{E}}
\newcommand{\var}{\text{var}}
\newcommand{\cov}{\text{cov}}
\newcommand{\cor}{\text{cor}}

\newcommand{\bX}{\mathbf{X}}

\newcommand{\bW}{\mathbf{W}}

\newcommand{\bP}{\mathbf{P}}

\newcommand{\Sc}{\mathcal{S}}

\newcommand{\indep}{\raisebox{0.05em}{\rotatebox[origin=c]{90}{$\models$}}}
\definecolor{shadecolor}{gray}{0.9}
\newcommand{\qed}{\hfill \ensuremath{\Box}}

\tikzset{every picture/.style={line width=0.75pt}} 

\usepackage{enumitem}
\newlist{Step}{enumerate}{2}
\setlist[Step]{label={{\textbf{Step \arabic*.}}}, leftmargin=*}

\usepackage{setspace}

\newcommand\blfootnote[1]{%
  \begingroup
  \renewcommand\thefootnote{}\footnote{#1}%
  \addtocounter{footnote}{-1}%
  \endgroup
}

 \makeatletter
\newcommand\circled[1]{%
  \mathpalette\@circled{#1}%
}
\newcommand\@circled[2]{%
  \tikz[baseline=(math.base)] \node[draw,circle,inner sep=2pt] (math) {$\m@th#1#2$};%
}
\makeatother

 \makeatletter
\newcommand\circledblue[1]{%
  \mathpalette\@circledblue{#1}%
}
\newcommand\@circledblue[2]{%
  \tikz[baseline=(math.base)] \node[draw,circle, fill=blue!20, inner sep=2pt] (math) {$\m@th#1#2$};%
 }

\renewenvironment{abstract}
 {\begin{center}\normalsize\textsc{Abstract}%
 \end{center}\begin{quote}\normalsize}
 {\end{quote}}

\title{Sensitivity Analysis for Survey Weights}
\author{Erin Hartman and Melody Huang}
\date{}

\usepackage[backend=biber,
style=authoryear,
natbib=true]{biblatex}
\begin{document}
\maketitle 
\begin{abstract}
Survey weighting allows researchers to account for bias in survey samples, due to unit nonresponse or convenience sampling, using measured demographic covariates. Unfortunately, in practice, it is impossible to know whether the estimated survey weights are sufficient to alleviate concerns about bias due to unobserved confounders or incorrect functional forms used in weighting. In the following paper, we propose two sensitivity analyses for the exclusion of important covariates: (1) a sensitivity analysis for partially observed confounders (i.e., variables measured across the survey sample, but not the target population), and (2) a sensitivity analysis for fully unobserved confounders (i.e., variables not measured in either the survey or the target population). We provide graphical and numerical summaries of the potential bias that arises from such confounders, and introduce a benchmarking approach that allows researchers to quantitatively reason about the sensitivity of their results. We demonstrate our proposed sensitivity analyses using state-level 2020 U.S. Presidential Election polls.
\end{abstract} 

\begin{center}
This Draft: \today
\end{center}

\blfootnote{The authors wish to thank Naoki Egami, 
Kirill Kalinin, Xiao Lu, and Erin Rossiter for thoughtful, valuable feedback.  We also wish to thank the participants of the UNC-Chapel Hill FAQ Seminar and the C-Dem Workshop: Survey Design Under Constraints--Best Practices.  Melody Huang is supported by the National Science Foundation Graduate Research Fellowship under Grant No. 2146752.}

\clearpage 
\doublespacing
\setcounter{page}{1}

\section{Introduction}
Given the fall of survey response rates 
and the increased reliance on online convenience samples, concerns about bias in surveys are rising. For example, \citet{bradley2021unrepresentative} discuss bias in public opinion surveys used to estimate COVID-19 vaccination rates.   Modern surveys nearly always rely on weighting adjustments, and researchers must carefully examine how survey weights are constructed.  For example, \citet{kennedy2018evaluation} found a primary driver of bias in the 2016 U.S. Presidential election was survey weights that failed to either account for education, or the interaction of education with race.  Subsequently, many surveys began including these when constructing survey weights.  Despite this, the 2020 U.S. Presidential election cycle resulted in national level public opinion polling that exhibited some of the most bias the last 40 years.  \citet{aapor2020} found the errors of 2016 do not explain problems in the 2020 election polls.  Instead, they find that bias is likely driven by unobservable confounders.

While these polling misses provide an opportunity to retrospectively evaluate these issues, they also make clear that within any specific survey, the question of how to properly construct survey weights is an open one.  

The purpose of our paper is to provide a set of sensitivity analyses that researchers can use on weighted surveys to analyze sensitivity to two types of confounders: (1) partially observed confounders, which are measured in the survey sample but not the target population, and (2) fully unobserved confounders which are not measured in the survey sample or the target population.  While these variables cannot be directly incorporated in the construction of weights, researchers are often aware of their existence through theoretical concerns.

In summary, we decompose bias in weighted estimators into two observable components--variation in the outcome and variation in the estimated weights--and two unobservable components--the correlation of the error in the estimated weights with the outcome, and how much variation in the ideal (true) weights this error explains.  We then propose two sensitivity analyses.  For partially observed confounders, our sensitivity analysis is based on a posited distribution of the confounder in the target population.  For fully observed confounders we propose a two parameter sensitivity analysis based on the unobserved components of our bias decomposition, and provide graphical and numerical summaries of robustness.

We extend the sensitivity analyses developed by \citet{huang2021saw}, \citet{hong2021did}, and \citet{shen2011sensitivity}, which assess the sensitivity in estimating causal effects with weighted estimators. Alternative approaches include sensitivity analyses which bound the worst-case bias from an unobserved confounder for matching estimators \citep[e.g.][]{rosenbaum1983assessing} and  marginal sensitivity models that consider a multiplicative error term in weighted estimators \citep[e.g.][]{soriano2021interpretable, zhao2019sensitivity, tan2006distributional}; and approaches that rely on invoking parametric or distributional assumptions about the underlying data generating processes \citep[e.g.][]{nguyen2017sensitivity}. 
An advantage of our framework is its generality; we do not require parametric assumptions on the data generating process for either the outcome or for sample selection. Our method is applicable to non-negative calibration weights and inverse propensity weights, making it relevant for many researchers.

\subsection{Running Example: 2020 U.S. Presidential Election}\label{subsec:running_example}
We consider a retrospective analysis of the 2020 U.S. Presidential election to demonstrate sensitivity analyses to partially and fully unobservable confounders. We evaluate two state-level ABC News/Washington Post (ABC/Wapo) polls conducted in October 2020. In particular, we examine polls for Michigan (October 20 - 25, 2020) and North Carolina (October 12 - 17, 2020). The ``survey samples'' consist of 770 respondents (Michigan) and 619 respondents  (North Carolina) who reported they were planning to vote or had already voted.

We want to demonstrate estimation of sensitivity parameters, including benchmarking against observed covariates, which requires we estimate custom weights rather than use the proprietary ones.  We construct weights calibrated to a target population defined using the 2020 Cooperative Election Study (CES) \citep{DVN/E9N6PH_2021}, following \citet{caughey_elements_2020}, which provides rich auxiliary data.\footnote{In practice, the CES, or any large survey that employs weighting against an administrative file or census used to define the target population, such as the GSS, Afrobarometer, or Eurobarometer, could conduct a sensitivity analysis.}  Our target population includes verified voters, incorporating the CES weights, \texttt{commonweight\_vv\_post}. %
We limit to units who stated they ``Definitely voted'' and for whom the retrospective candidate choice for the 2020 Presidential Election is not missing. This defines our ``target population'' and provides estimates of candidate support at the state level as a point of reference for our analysis. Using this approach, we have 1,213 units in our ``target population'' for Michigan, and 1,101 units for North Carolina.  

We construct calibration weights by raking which, in short, ensures the survey sample is representative by matching the weighted survey mean of observed demographic characteristics to known target population means. 
We include the following covariates in the construction of our survey weights: age, gender, race/ethnicity, educational attainment, party identification, and an indicator for born-again Christian.\footnote{Whether or not to weight on party identification is an open debate.  In Supplementary Materials~E we demonstrate our sensitivity analysis for Michigan without weighting on party.}  For simplicity, we conduct raking only on marginal population means, as well as a two-way interaction between party identification and educational attainment, but alternative methods can account for higher order interactions \citep[e.g.][]{hartman2021kpop, DRP}.

\begin{table}[ht]
\centering
\begin{tabular}{lccccc}
  \toprule
State & Unweighted & Weighted & CES & True Election Result\\ 
  \midrule
Michigan & 7.25 (3.59) & 4.57 (2.56) & 2.06 & 2.78 \\ 
  North Carolina & 0.65 (4.04) & -0.37 (2.63) & -5.10 & -1.35 \\ 
   \bottomrule
\end{tabular}
\caption{Unweighted and Weighted Margin (D-R) in two-party vote share in percentage points. We provide the CES estimate and the true vote margin as reference.
} 
\label{tab:weighted_app} 
\end{table}

Our primary outcome of interest is the Democratic margin in the two-party vote share (Democrat (D) - Republican (R)) among those who state a preference for either major party candidate. Table~\ref{tab:weighted_app} presents the unweighted and weighted outcomes.  The goal of our method is to determine how sensitive the weighted point estimate is to partially or fully unobserved confounders. 
As is common in U.S. presidential elections, the election is predicted to be close and the estimates are statistically indistinguishable from zero, however the point estimate provides the best prediction for the substantive outcome; thus we focus on sensitivity in the point estimate. We describe how to incorporate sensitivity in measures of statistical uncertainty in the conclusion.  Our sensitivity analysis tools allow researchers to transparently reason about whether their findings---in this case who leads in each state---are robust to the exclusion of confounders when estimating weights.

\section{Notation and Set-up}\label{sec:notation}
We consider a finite target population with $N$ units. The survey sample consisting of $n$ units is drawn from the target population (where $n << N$); we assume the survey sample is not a simple random sample, and as such, is not representative of the target population. Let $S_i \in \{0,1\}$ be a survey inclusion indicator, where $S_i = 1$ when a unit is a respondent in the survey and 0 otherwise.  Let $\Pr(S_i = 1)$ denote the probability of inclusion for unit $i$.\footnote{While the current set-up assumes that the survey sample is a subset of the target population (`nested'), our results extend to a non-nested set-up. See \citet{hartman2021kpop}, \citet{huang2021leveraging}, and \citet{huang2021saw} for discussion.}  Note that for a probability sample without nonresponse, $\Pr(S_i = 1)$ represents the sampling probability; for a convenience sample or a sample with nonresponse, it encodes the product of the sampling probability and response probability for unit $i$, both of which may be unknown.  Throughout the manuscript, we refer to the collection of respondents, or units for which $S_i = 1$, as the ``survey sample'' (denoted $\mathcal{S}$) and we assume that each unit in the target population has some positive (but possibly unknown) probability of inclusion in the survey sample (i.e. $0 < \Pr(S_i = 1) \le 1$).  We denote quantities calculated over the survey sample with a subscript $\mathcal{S}$ (e.g. $\var_\mathcal{S}(\cdot)$).  Our outcome of interest is $Y$, which is observed for every unit in the survey sample.  Our target estimand is the population mean $\mu = \frac{1}{N} \sum_{i = 1}^{N} Y_i$.  See Table~A.1 in the Supplementary Materials for a glossary of terms.

When the survey sample is a simple random sample from the target population, then the average outcome within the survey sample is an unbiased estimator for the population mean. However, in most cases, the survey suffers from unit nonresponse, or is a convenience sample, resulting in a non-representative sample; as such, using the mean outcome within the survey sample may be a biased estimator of the population mean.  We assume that to account for non-random selection into the survey sample, researchers construct survey weights that adjust the survey to be representative of the target population on observable characteristics.

One way to construct survey weights is to assume that the observable characteristics $\bX$ are sufficient, and conditional on these variables the distribution of the outcome among the survey respondents and the target population is the same.

\begin{assumption}[Conditional Ignorability of Response \citep{little2019statistical}]  \label{assump:cond_ig} 
    $$Y \ \indep \ S  \mid \bX$$
\end{assumption} 

Assumption~\ref{assump:cond_ig} non-parametrically justifies post-stratification weights constructed by dividing the population proportion by the survey proportion within intersectional strata defined by $\bX$.  Common approaches include conditioning on variables that fully explain sampling or the outcome, but many alternatives are possible \citep{egami2019covariate}. 

When weighting on continuous variables or a large number of strata, post-stratification suffers from sparsity constraints.  In this case, researchers must define a feature mapping $\bX \mapsto \phi(\bX)$ from $\mathbb{R}^P \mapsto \mathbb{R}^{P'}$ that captures important features of $\bX$ for use in coarsened post-stratification or model-based weighting methods. Researchers can construct calibration weights, subject to a set of moment constraints defined on $\phi(\bX)$.  For example, researchers can use ``raking'' to calibrate the marginal means on all variables $\bX$, letting $\phi(\bX) = \bX$.  Alternatively, inverse propensity weights (IPW), in which researchers directly estimate probability of inclusion and weight inversely proportional to this estimate may use $\phi(\bX) = \bX$, which are included link-linearly in logistic regression.  We focus our conceptual discussion around IPW, which are asymptotically equivalent to calibration weights \citep{
ben2021balancing} but use calibration for estimation.  In particular, we employ raking on the margins, a type of calibrated IPW, in our analysis.  A detailed discussion on the construction of survey weights is beyond the scope of this paper.  See Supplementary Materials~B.1 for a discussion of calibration weights; we refer readers to \citet{haziza_beaumont_2017} 
for a thorough review on how to construct survey weights and related considerations.

How to construct an appropriate feature mapping is an active field of research.  For example, recent methods aim to include higher-order interactions of $\bX$ \citep[e.g.][]{DRP} or rely on kernel methods to account for important features \citep[e.g.][]{hartman2021kpop}.  While these methods are important for flexibly accounting for observable characteristics, our proposed sensitivity analysis also evaluates robustness to unobservable confounders.

When relying on model-based weighting adjustments, such as calibration, researchers may not be able to directly invoke Assumption~\ref{assump:cond_ig}.  Instead, they can appeal to a linear ignorability assumption for consistent estimation of the population mean:

\begin{assumption}[Linear ignorability in $\phi(\bX)$ \citep{hartman2021kpop}]\label{assump:linearign}
Let $Y_i=\phi(\bX_i)^{\top}\beta + \delta_i$ and $Pr(S_i=1|\bX_i) = g(\phi(\bX_i)^{\top}\theta + \eta_i)$ where $g(\cdot):\mathcal{R}\mapsto [0,1]$. Linear ignorability holds when $\delta_i \indep \eta_i$.
\end{assumption}

\noindent Assumption~\ref{assump:linearign} states that the part of $Y$ orthogonal to $\phi(\bX)$ must be independent of the part of $S$, the survey selection process, orthogonal to $\phi(\bX)$ via a suitable link function.  See \citet{hartman2021kpop} for more details.  What Assumption~\ref{assump:linearign} makes clear is that researchers must carefully choose a feature mapping, $\phi(\bX)$ to account for all features of $\bX$, including interactions and transformations, that affect both survey inclusion and the outcome.

Assumption~\ref{assump:cond_ig} and Assumption~\ref{assump:linearign} are related.  The advantage of Assumption~\ref{assump:linearign} is that it allows researchers to address sparsity in finite data by imposing a parametric assumption, whereas Assumption~\ref{assump:cond_ig} nonparametrically identifies the population mean.  They differ most starkly in the types of violations that are problematic. 
Given a set of survey weights constructed using feature mapping $\phi(\cdot)$, we define the weighted estimator as follows:

\begin{definition}[Weighted Estimator for Population Mean]
$$\quad \hat \mu = \sum_{i \in \Sc} w_i Y_i.$$
\end{definition}

\noindent This estimator could be biased for two reasons: (1) there is an unobserved confounder $U$ 
(violating Assumptions~\ref{assump:cond_ig}~and~\ref{assump:linearign}), or (2) researchers have failed to construct an adequately rich feature expansion $\phi(\bX)$ of the observable characteristics, such as by failing to incorporate higher-order moments or interactions (violating Assumption~\ref{assump:linearign}). These ignorability assumptions are strong and untestable, although some observable implications can be tested. While survey weights might make the sample representative on the features included in estimation, this does not imply that these ignorability assumptions hold. Weighting will typically mitigate bias if the weighting variables are correlated with the outcome, although it could exacerbate bias.  As we can never know if Assumption~\ref{assump:cond_ig} or \ref{assump:linearign} hold, it is essential we have tools to transparently evaluate their credibility. The focus of our paper is to help researchers evaluate if their results are sensitive to the exclusion of unobservable characteristics or more complex observable features.

\section{Bias in Weighted Estimators}\label{sec:bias}

In the following section, we introduce the bias of weighted estimators when omitting an unobservable variable $U$ from estimation of the survey weights.  More specifically, under Assumption~\ref{assump:linearign}, we assume that $Y$ can be decomposed into $\phi(\bX)^\top\beta + U + \nu$, where $\nu \indep \{X, U, S\}$. Implicitly, this means $U$ is included in $\delta = U + \nu$.  Moreover, we assume $\beta$ and $\theta$ are non-zero, indicating there is a correlation between $Y$ and $S$, weighting is necessary, and that the true weights are not all equal to 1.  This would be violated, for example, if the survey was a simple random sample with no nonresponse.

The omitted variable set-up provides a flexible formulation for assessing violations of Assumption~\ref{assump:linearign}.  Without loss of generality, we assume $U$ is orthogonal to $\phi(\bX)$, and could be a combination of multiple underlying variables.  $U$ can always be replaced by the residual from projecting confounders onto $\phi(\bX)$. 
This is important when reasoning about potential confounders; it makes clear that bias is due to the part of the omitted confounder not linearly explained by the included covariates.  For example, if we have omitted political interest, but we have included age and educational attainment in the construction of our weights, $U$ is the part of political interest not linearly explained by age and educational attainment.  This indicates that potential confounders well explained by $\phi(\bX)$ are less problematic.  If $U$ is associated with selection into the survey, i.e. correlated with $S$, then Assumption~\ref{assump:linearign} is violated because $\delta_i \not\!\perp\!\!\!\perp \eta_i$. The goal of the sensitivity analysis is to assess the robustness of point estimates to the omitted confounder $U$.

Alternatively, $U$ could be interactions or non-linear transformations of observable covariates $\bX$ not included in $\phi(\bX)$. For example, \citet{kennedy2018evaluation} showed that simply accounting for the linear combination of race/ethnicity, region, and educational attainment is insufficient; instead, researchers also needed to account for an interaction between race/ethnicity, region, and educational attainment.

\subsection{Derivation} 
We begin by defining the vector of estimated weights, $w$, as those estimated using $\phi(\bX)$, and the vector of ideal weights $w^*$ as those that would have been estimated including $\phi(\bX)$ and $U$. 
Throughout the paper, we will assume, without loss of generality, that both the estimated and ideal weights are centered at mean 1. Finally, define the error term $\varepsilon$ as the difference between the estimated weights and the ideal weights (i.e., $\varepsilon := w - w^*$).

The error in the weights is driven by the \textit{residual} imbalance in an omitted confounder, after balancing on 
$\phi(\bX)$.
This general intuition holds regardless of the weighting approach: if imbalance in $U$ is minimal after adjusting for $\phi(\bX)$, the overall error from omitting $U$ when constructing weights should be low.  In Supplementary Materials~C.1 we derive the error for IPW weights, highlighting where imbalance factors into the error.  In the context of our running example, an individual's baseline level of political interest is an important predictor of both survey inclusion and many political outcomes; however it is not typically incorporated into weight estimation because there are no population measures. If political interest is largely explained by variables included in weight estimation, such as age and educational attainment, then the residual imbalance in political interest should be small, and $\varepsilon_i$ should be relatively small.

The bias in $\hat \mu$ from omitting the variable $U$ from estimation of the survey weights can be parameterized with respect to the error term, $\varepsilon$. We formalize this in the following theorem. 

\begin{theorem}[Bias of a Weighted Estimator]
Let $w$ be the weights estimated using just $\phi(\bX)$ and $w^*$ be the ideal weights estimated using $\phi(\bX)$ and $U$. The bias in $\hat \mu$ from omitting $U$ from estimation of the weights can be written as: 
\label{thm:bias} 
\begin{align}
    \text{Bias}(\hat \mu) &=  \E(\hat \mu) - \mu \nonumber \\
    &= \begin{cases} \displaystyle 
    \textcolor{blue}{\rho_{\varepsilon, Y}} \sqrt{\var_\Sc(Y) \cdot \var_\Sc(w) \cdot \frac{\textcolor{blue}{R^2_\varepsilon}}{1-\textcolor{blue}{R^2_\varepsilon}}} &\text{if }  R^2_\varepsilon < 1 \\
    \displaystyle \textcolor{blue}{\rho_{\varepsilon, Y}} \sqrt{\var_\Sc(Y) \cdot \var_\Sc(w^*)} &\text{if } R^2_\varepsilon = 1
    \end{cases} 
    \label{eqn:bias_decomp},
\end{align}
where $\varepsilon$ is the error in the weights from omitting $U$, $R^2_\varepsilon$ is the ratio of variation in the ideal weights explained by $\varepsilon$ (i.e., $R^2_\varepsilon := \var_\Sc(\varepsilon)/\var_\Sc(w^*)$), and $\rho_{\varepsilon, Y}$ is the correlation, or alignment, between $\varepsilon$ and the outcome $Y$ (i.e., $\rho_{\varepsilon, Y} := \cor_{\Sc}(\varepsilon, Y)$). Quantities denoted with a subscript $\Sc$ are estimated over the survey sample (i.e., $S_i = 1$). Proof in Supplementary Materials~3.1.
\end{theorem} 
Theorem \ref{thm:bias} decomposes the sources of bias from omitting a confounder in weight estimation. The terms highlighted in blue are unobserved, while the other terms in black can be directly estimated from the survey data using sample analogs.  By treating the weights as fixed, we implicitly derive the asymptotic bias.  See Supplementary Materials~3.1 for a discussion of the finite-sample case. We provide more details about the interpretation and properties of each component of the bias formula in the following subsection.

\subsection{Interpreting the Drivers of Bias}
In the following subsection, we discuss the different components in the bias formula from Theorem \ref{thm:bias}: (1) $R^2_\varepsilon$, representing the ratio of the variance in the error $\varepsilon$ and the variance in the ideal weights, (2) $\rho_{\varepsilon, Y}$, the alignment between the error in the weights, $\varepsilon$, and the outcome $Y$, and (3) a scaling factor (i.e., $\var_\Sc(Y) \cdot \var_\Sc(w)$). The first two components are unobservable, while the last component is directly estimable from the observed data. We show that both of the unobserved components exist on bounded, standardized ranges, and provide intuition for how to interpret each component. 

\subsubsection{Explained variation in ideal weights ($R^2_\varepsilon$)}
$R^2_\varepsilon$ represents the amount of variation in the ideal survey weights $w^*$ that is explained by the error term $\varepsilon$. Following \citet{huang2021saw}, we decompose the total variation in the ideal weights into two components: (1) the amount of variation in $w^*$ explained by the estimated weights $w$, and (2) the amount of variation in the ideal weights $w^*$ explained by the estimation error $\varepsilon$: 

\begin{corollary}[Variance Decomposition of $w^*$ (\citet{huang2021saw})] \label{cor:var_decomp} 
Let $w$ be estimated IPW and $w^*$ be the ideal weights. The variance of the ideal weights $w^*$ can be decomposed linearly into two components: 
\begin{align*} 
\var_\Sc(w^*) &= \var_\Sc(w) + \var_\Sc(\varepsilon)
\implies \frac{\var_\Sc(w)}{\var_\Sc(w^*)} + \underbrace{\frac{\var_\Sc(\varepsilon)}{\var_\Sc(w^*)}}_{:= R^2_\varepsilon} = 1
\end{align*} 
\end{corollary}
An implication of Corollary \ref{cor:var_decomp} is that $R^2_\varepsilon$ is guaranteed to be bounded on the interval $[0,1]$.

As the residual imbalance in the omitted confounder $U$ increases, $R^2_\varepsilon$ increases.  Intuitively, if imbalance in $U$ across the target population and survey sample is large, the ideal weights $w^*$ will be very different from the estimated weights $w$. As a result, the variance in the estimation error, $\var_\Sc(\varepsilon)$, will be large, thus increasing $R^2_\varepsilon$. In contrast, if residual balance in $U$ is small, the error will be small and thus the ideal weights $w^*$ will be close to the estimated weights, leading to small $R^2_\varepsilon$.\footnote{In the scenario that $R^2_\varepsilon = 1$ (i.e., the error from omitting the confounder can explain 100\% of the variation in the ideal weights $w^*$), researchers will need to posit values for $\var_\Sc(w^*)$ to estimate the bias.%
This occurs, for example, with no weighting adjustment and unity weights.  However, we argue that if researchers have conducted weighting and accounted for even a single covariate that can explain any variation in the selection process, $R^2_\varepsilon < 1$.}

Consider our running example, in which retrospective studies showed that voters with lower levels of educational attainment were underrepresented in the survey sample, relative to the target population. State polls, particularly in the Midwest, that omitted educational attainment, despite accounting for other demographic variables, saw significant error in point estimates.  If educational attainment is not well explained by the other demographic characteristics, this will lead to a large $R^2_\varepsilon$.  The bias is exacerbated by the correlation between this error and the outcome, since white voters with lower educational attainment were more likely to support Donald Trump in the 2020 U.S. Presidential election.

\subsubsection{Alignment between the error and outcome ($\rho_{\epsilon, Y}$)} 
The alignment\footnote{We use the term ``alignment'' based on a similar concept in \citet{kern2016assessing}.} between the error in the weights and the outcome, captured by the correlation $\rho_{\epsilon, Y}$, also affects the potential for bias. For example, if positive $\varepsilon$ values correspond to large $Y$ values, then this implies that units that are being over-represented also have a larger $Y$ value, resulting in positive bias. Conversely, if negative $\varepsilon$ values correspond to large $Y$ values, this implies units with a large $Y$ value are under-represented, resulting in negative bias.

In our running example, assume the survey sample has higher average baseline political interest than the target population and also that, conditioning on our weighting variables, individuals with greater political interest are more likely to vote for a Democratic candidate.  Omitting political interest from weight estimation would lead to larger weights placed on Democratic voters in the survey, resulting in a bias that overstates Democratic vote share.

\subsubsection{Bias Scaling Factor} 
The final terms in Equation \ref{eqn:bias_decomp}, $\var_\Sc(w)$ and $\var_\Sc(Y)$, are not dependent on the unmeasured confounder $U$ and are directly estimable from the survey data.  They act as a scaling factor that increases the sensitivity of the point estimate to violations of the assumptions, even if the alignment $\rho_{\varepsilon, Y}$ and the variance explained by the error $R^2_\varepsilon$ are close to zero.

It is well known that if $\var_\Sc(w)$ is large, such as from extreme weights, this increases the variance in the estimator.  This is related to the design effect \citep{kish1965}, and underscores the importance of choosing prognostic and substantively meaningful variables to weight on as well as design-stage considerations that minimize the variance in the weights.  This also emphasizes the bias-variance trade-off when constructing survey weights.

The term $\var_\Sc(Y)$ is the variance of the outcome variable $Y$, related to the ``problem difficulty'' in \citet{meng2018statistical}, which is not something that researchers can control. It formalizes the intuition that if the outcome variable $Y$ has a high degree of heterogeneity it is potentially more sensitive to sources of bias.

The bias decomposition highlights an important point: an omitted variable must be related to both the outcome \textit{and} the response process in order for there to be bias from omitting it. Even if the omitted variable is imbalanced between the survey sample and the target population (i.e., $R^2_\varepsilon > 0$), if it is not related to the outcome (i.e., $\rho_{\varepsilon, Y} = 0$), there will be no bias. Similarly, if an omitted variable is related to the outcome (i.e., $|\rho_{\varepsilon, Y}| > 0$), but is balanced between the survey sample and the target population (i.e., $R^2_\varepsilon = 0$), no bias will occur. This framework helps formalize the types of variables that researchers should consider when assessing the sensitivity in their estimates. 

\subsection{Performing Sensitivity Analyses} 
The bias decomposition in Theorem \ref{thm:bias} provides a natural basis for performing a sensitivity analysis. By positing values for the unobserved parameters, $R^2_\varepsilon$ and $\rho_{\varepsilon, Y}$, researchers can estimate bias and evaluate the robustness to residual confounding. In the following sections, we propose two sensitivity analyses. The first sensitivity analysis (Section \ref{sec:partial_observe}) shows that when a confounder is observed across the survey sample but not the target population, Theorem~\ref{thm:bias} can be re-written as a function of a single unobserved parameter. The second sensitivity analysis (Section \ref{sec:unobserve}) allows researchers to assess the sensitivity to fully unobserved confounders using a two parameter sensitivity analysis. We propose a set of tools that allow researchers to (1) summarize the amount of sensitivity in their point estimate, and (2) benchmark the analysis using observed covariate data.

\section{Sensitivity Analysis for Partially Observed Confounders} \label{sec:partial_observe} 
Researchers typically have greater control over the variables they can measure among survey respondents, and they may have strong theoretical reasons to believe a variable is related to both sampling and the outcome; however, if the covariate is not measured across the target population, it cannot be incorporated into the weights and thus it should be assessed in sensitivity analyses. We formalize this issue by defining a variable measured in the survey sample, but not in the target population, as a \textit{partially observed confounder}, denoted as $V$.  This extends to excluded functions of observables, for example higher-order moments or interactions of the observable covariates. 

In this section we propose a sensitivity analysis that evaluates robustness to partially observed confounders against a hypothetical distribution of this confounder in the target population.  These confounders can be identified using theory, but in Supplementary Materials~D.2, we suggest a data-driven approach for detecting such confounders.  This is useful in settings when researchers are unsure from a purely substantive standpoint whether or not a partially observed variable must be included in the weights.  A summary of our suggested approach for detecting and evaluating sensitivity to partially unobserved confounders is provided in Figure~\ref{fig:sens_partially}.

Recall that the error in the weights is driven by the residual imbalance in an omitted confounder (see Supplementary Materials~C.1 for details). With information about $V$ across the survey sample, the sensitivity analysis can be reduced to one sensitivity parameter: the distribution of $V$ in the target population.  We focus here on how to incorporate such a parameter in calibration weighting, and discuss additional details for IPW in Supplementary Materials~D.1.

With calibration weighting, raking on the margins, a sensitivity analysis for partial confounding reduces to including an additional moment constraint for $\E(V)$, or the posited population mean, across a range of plausible values for the population mean.  Calibration will solve for weights that will simultaneously meet the original and the additional moment constraints, even without knowing the joint distribution of the original covariates with the partially observed covariate. The resulting error in the weights can be directly calculated and used to estimate bias.  When $V$ is binary, $\E(V)$ is bounded on $[0,1]$; this is easy to extend to categorical variables with three levels, but difficult to visually evaluate beyond that.  A limitation to this approach is that when $\E(V)$ is unbounded, researchers must specify a theoretically relevant range for $\E(V)$, informed through substantive knowledge or existing data on $\E(V)$ from similar target populations.  If researchers do not have a strong substantive prior for a reasonable range of $\E(V)$, it may be helpful to redefine $\E(V)$ in terms of standard deviations from the sample average (see \citet{nguyen2017sensitivity} for more discussion). This naturally extends when raking across additional moments or interactions, although it requires the researcher to specify more parameters.

\begin{figure}
\caption{\underline{\textbf{Sensitivity Analysis for Partially Observed Confounders}}\label{fig:sens_partially}}
\noindent\fbox{%
\vspace{2mm}
\parbox{0.98\textwidth}{%
\vspace{2mm}
\begin{Step} 
\item Use theory or data-driven method to detect $V$ \citep{egami2019covariate}, see Supplementary Materials~D.2.
\item Posit values for the population average of $V$ (i.e., $\E(V_i \mid \bX_i)$ for propensity score weights, or $\E(V)$ for calibration weights) and re-estimate the survey weights. 
\item Use the re-estimated weights to obtain an updated $\hat \mu$.
\end{Step} 
}
}\\
\end{figure}

\subsubsection{Running Example: 2020 U.S. Presidential Election} 

Existing literature indicates that political participation is also correlated with propensity to respond to surveys \citep{peress2010correcting}; however it is not commonly incorporated into survey weights because it is not available in target populations data. To illustrate our proposed sensitivity analysis for partially observed confounders, we focus on the Michigan poll. We use interest in the upcoming election as a proxy for political participation, and posit it is a partially observed confounder.  Below, to visually demonstrate sensitivity analysis for a partially observed confounder, we re-code this as a binary variable.\footnote{We define political interest `1' for individuals who responded that they were following the 2020 election ``very closely'', and `0' otherwise.}

\begin{figure}[!ht] 
\centering
\includegraphics[width=0.8\textwidth]{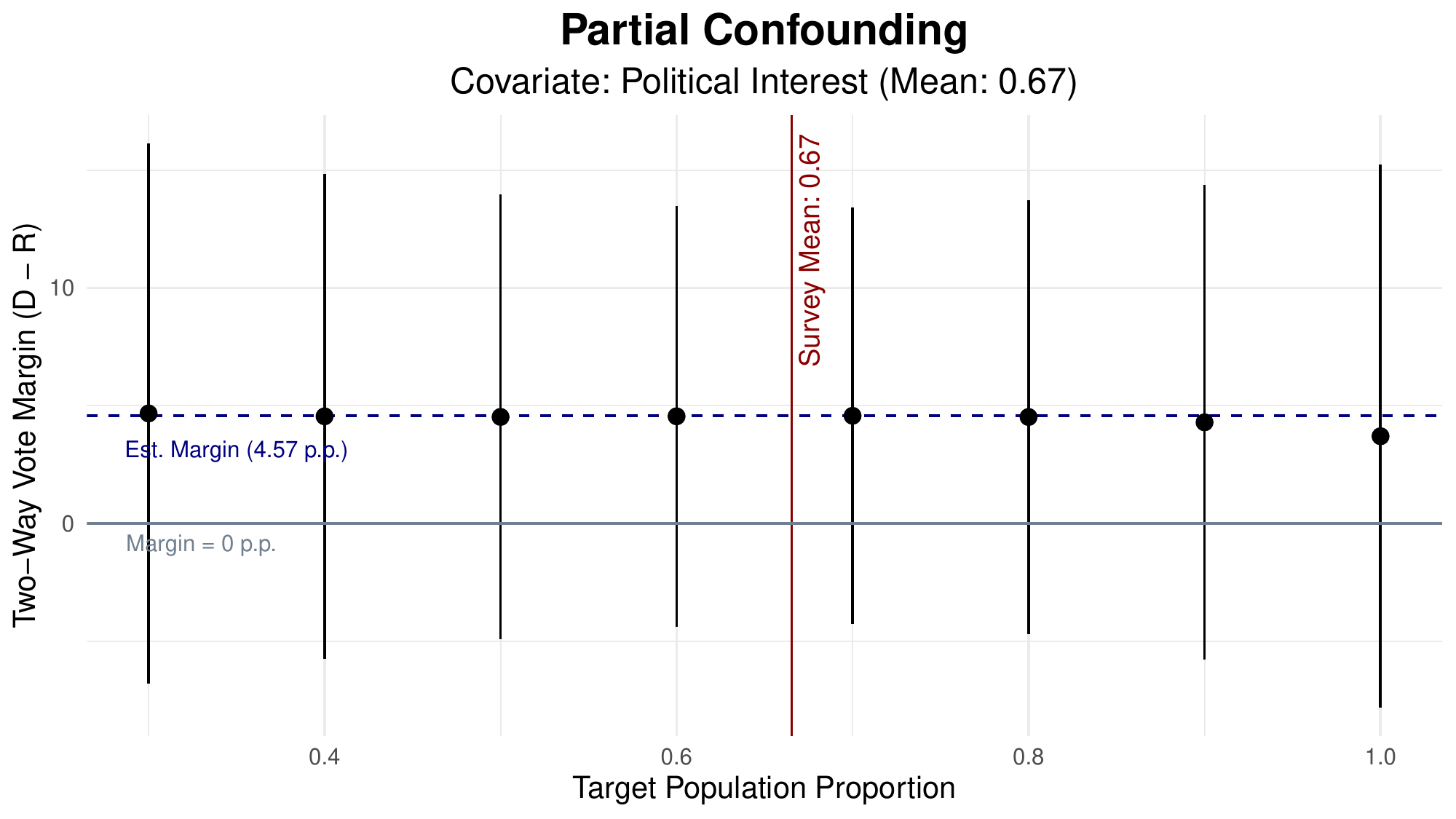}
\caption{Plot of estimates for two-way vote margin (D-R) as the proportion of politically interested individuals in the target population changes.
} 
\label{fig:partial_confounding} 
\end{figure}

In the survey, 67\% of the respondents are encoded as ``very closely'' following the 2020 election, using the estimated weights. The sensitivity analysis varies the proportion of individuals in the target population who are politically interested, from 0\% to 100\%, the natural range of the variable, and re-estimates the weights and evaluates how the point estimate changes. (See Figure \ref{fig:partial_confounding} for visualization.) The range could be reduced with a strong substantive argument or external data.  We see that the point estimate is very insensitive to excluding political interest, moving very little across the entire range of the sensitivity parameter, and that the substantive result (i.e., Democratic two-way vote share is greater than 0) is robust to omitting this partially observed confounder.  Standard errors include 0 across the full range of the sensitivity parameter.  We argue that the analysis is very insensitive to this partially observed confounder, and researchers need not include it in the weights in this survey.  In Supplementary Materials~D.2.1, we confirm that this political interest variable is, in fact, not a partially observed confounder using our algorithm for detecting such confounders; political interest is rendered irrelevant to the outcome using only fully observed covariates within the survey sample.

\section{Sensitivity Analysis for Fully Unobserved Confounders} \label{sec:unobserve}

In the following section, we introduce a sensitivity analysis for fully unobserved confounders. The method uses Theorem~\ref{thm:bias} as the foundation for a two parameter sensitivity analysis. Researchers posit values for both the alignment, $\rho_{\varepsilon, Y}$, and the variation in the ideal weights explained by the error, $R^2_\varepsilon$, to evaluate potential bias in $\hat \mu$. We summarize our approach in Figure~\ref{fig:sens_unobserved}.

\begin{figure}
    \caption{\underline{\textbf{Summary: Sensitivity Analysis for Unobserved Confounders}} \label{fig:sens_unobserved}}
\noindent\fbox{%
\vspace{2mm}
\parbox{0.98\textwidth}{%
\vspace{2mm}
\begin{Step} 
\item Using the observed survey sample data, estimate the variance of the outcome $Y$ and the estimated weights $w$ (i.e., estimate $\var_\Sc(Y)$ and $\var_\Sc(w)$).
\item Vary $\rho_{\varepsilon, Y}$ across the range $[-1,1]$. 
\item Vary $R^2_\varepsilon$ across the range $[0,1)$.
\item Evaluate the bias of $\hat \mu$ for each $\rho_{\varepsilon, Y}$ and $R^2_\varepsilon$ value using Equation~\ref{eqn:bias_decomp}. 
\item Interpret bias using the robustness value $RV_{b^*}$, contour plots, and benchmarking against observables.
\end{Step} 
}
}\\
\end{figure}

To help researchers conduct the sensitivity analysis, we propose three methods that allow researchers to (1) summarize the degree of robustness in their point estimate using a single ``robustness value'', (2) graphically evaluate robustness to fully unobserved confounders with bias contour plots, and (3) benchmark potential bias using observed covariates.

\subsection{Numerical Summary: Robustness Value} 
We propose a standardized numerical summary of sensitivity, in the form of a ``robustness value'', $RV_{b^*}$, to allow researchers to succinctly assess the plausible existence of fully unobserved confounders.  Researchers must first specify a substantively meaningful threshold ${b^*}$ at which bias would change the substantive conclusion.  The robustness value then represents the minimum amount of variation that the error term $\varepsilon$ must explain in both the ideal weights $w^*$ and the outcome $Y$ in order for the bias to be large enough to change the substantive result.\footnote{When $R^2_\varepsilon = \rho_{\varepsilon, Y}^2 = RV_{b^*}$, then the bias will equal $\hat \mu - {b^*}$.}  Following \citet{huang2021saw} and \citet{cinelli2020making}, the robustness value is estimated as:  
\begin{equation} 
RV_{b^*} = \frac{1}{2} \left( \sqrt{a_{b^*}^2 + 4a_{b^*}} - a_{b^*} \right), \ \ \ \text{where } a_{b^*} = \frac{(\hat \mu - {b^*})^2}{\var_\Sc(Y) \cdot \var_\Sc(w)}.
\label{eqn:RV} 
\end{equation} 

Researchers can pick different target values, ${b^*}$, that are substantively meaningful; for example, in the context of U.S. election polling, if researchers are interested in estimating a candidate's vote share, a logical value is ${b^*} = 0.5$ which represents the threshold for where candidate's vote share changes past the 50\% threshold, thus changing the prediction that the candidate will win or lose.  Similarly, if the outcome is vote margin, a logical value is $b^* = 0$, which indicates where the predicted winning candidate would change.  Other natural $b^*$ also include substantively meaningful deviations from the point estimate, such as a 20\% difference.  In the conclusion, we discuss how to additionally incorporate uncertainty in the estimation of weights using a percentile bootstrap to determine the bias necessary to change the statistical significance of the results \citep{huang2022variance}.

$RV_{b^*}$ is bounded on an interval from 0 to 1. When $RV_{b^*}$ is close to 1, this implies that error in the weights must explain close to 100\% of the variation in both $w^*$ and $Y$ in order for the bias to substantively change the point estimate. On the other hand, when $RV_{b^*}$ is close to 0, then if the error in the weights can account for even a small amount of variation in $w^*$ and $Y$, the resulting bias will be large enough to alter the substantive result of the point estimate.

\subsubsection{Running Example: 2020 U.S. Presidential Election}

Recall from Section~\ref{subsec:running_example} that the point estimate for our weighted ABC/Wapo poll projects Biden to win the popular vote by a 4.57 p.p margin ($\pm$ 2.56 p.p) in Michigan, and Biden to lose the popular vote by a -0.37 p.p. margin ($\pm$ 2.63 p.p.) in North Carolina.  We let ${b^*} = 0$, indicating we are interested in bias that would change the predicted winning candidate.  For the Michigan poll, $RV_{{b^*} = 0} = 0.11$; the error in the weights needs to explain 11\% of the variation in both the outcome and the ideal weights to reduce the estimated vote margin to zero.  Researchers need to substantively defend if this is plausible.  For the North Carolina poll, $RV_{b^* = 0} = 0.01$; the error only needs to explain 1\% of the variation in both the outcome and the ideal weights to reduce the estimated vote margin to zero. As such, we conclude that there is a greater degree of sensitivity in the North Carolina poll to an omitted variable substantively altering the predicted winner. 

\begin{table}
\centering
\begin{tabular}{lccc} \toprule 
 & Point Estimate & Standard Error & $RV_{{b^*} = 0}$ \\ \midrule 
Michigan & 4.57 p.p. & 2.56 p.p. & 0.11 \\ 
  North Carolina & -0.37 p.p. & 2.63 p.p. & 0.01 \\ 
  \bottomrule
\end{tabular} 
\caption{Point Estimate and Robustness Value for ABC/Wapo 2020 U.S. Presidential Election Poll.} 
\label{tbl:summary} 
\end{table} 

\subsection{Graphical Summary: Bias Contour Plots} 
While the robustness value is a useful summary measure,  $RV_{b^*}$ only represents a single point of the combination of $\{\rho_{\varepsilon, Y}, R^2_\varepsilon\}$ that could lead to substantively meaningful bias.  However, the estimation error may not equally explain the alignment and the variation in the ideal weights. 

To provide a fuller understanding of how bias may vary across different $\{\rho_{\varepsilon, Y}, R^2_\varepsilon\}$ values, we propose the use of bias contour plots. To construct the bias contour plots, researchers evaluate the bias at values of $\rho_{\varepsilon, Y}$ in the range $[-1,1]$ on the $x$-axis and $R^2_\varepsilon$ in the range $[0,1)$, on the $y$-axis. The bias is calculated using Equation~\ref{eqn:bias_decomp}.  This approach fully captures potential bias across the range of both of the sensitivity parameters.

The contour plots allow researchers to visualize the ``killer confounder'' region, which represents the values of $\{\rho_{\varepsilon, Y}, R^2_\varepsilon\}$ for which the bias is large enough to substantively change the meaning of the point estimate. The boundary of the killer confounder region is defined by $b^*$ (i.e., the same threshold value chosen for the robustness value); $RV_{b^*}$ is one point on the boundary of the killer confounder region, where $\rho_{\varepsilon, Y} = R^2_\varepsilon$. As such, it is important to report bias contour plots in order to assess the full set of possible parameter values that may result in a killer confounder. 

If the area of the killer confounder region dominates much of the plot, then the point estimate is very sensitive to fully unobserved confounders; in contrast, if the area is relatively small and contained to regions that are defensible as unlikely, then the result is robust.  

\subsubsection{Running Example: 2020 U.S. Presidential Election}

\begin{figure}[!ht]
\centering
\includegraphics[width=0.48\textwidth]{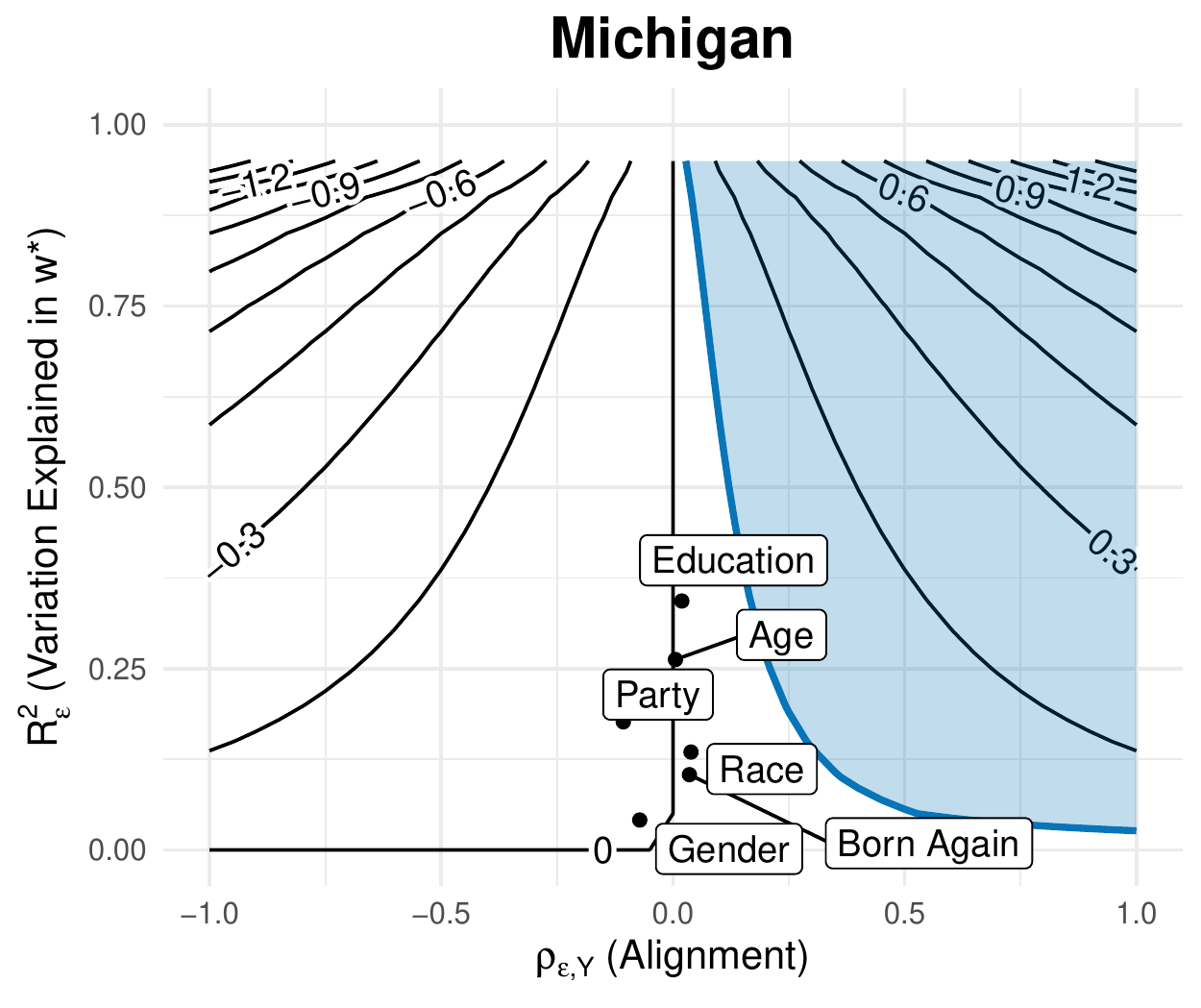}
\includegraphics[width=0.48\textwidth]{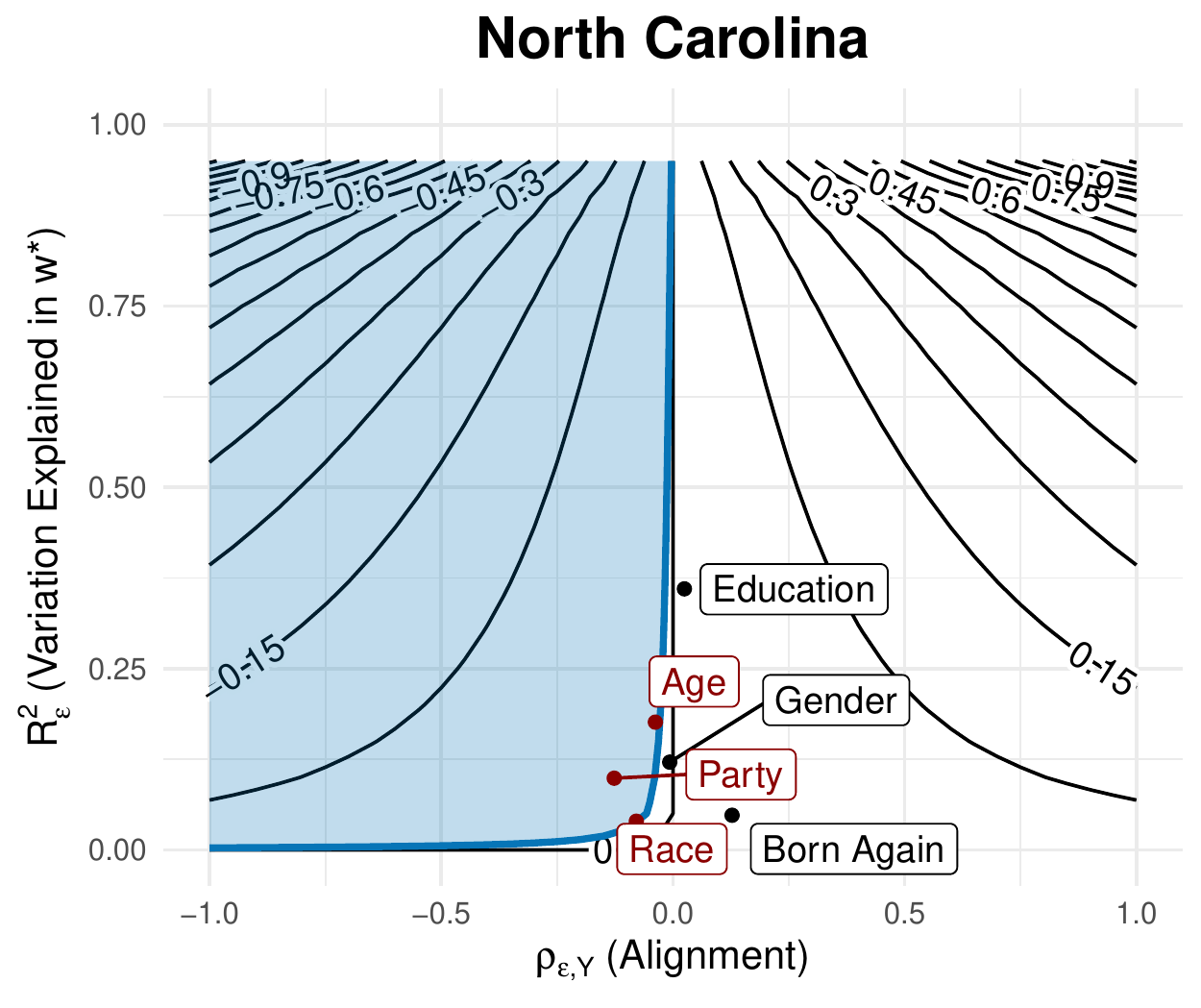}
\caption{Bias contour plots. The shaded blue region represents the killer confounder region, in which the bias is large enough to reduce the margin to or below zero, changing the predicted winner. We also plot the results from formal benchmarking, where each point represents the parameter value of an omitted confounder with equivalent confounding strength of an observed covariate.} 
\label{fig:bias_contour} 
\end{figure} 

Figure~\ref{fig:bias_contour} presents the bias contour plot for the ABC/Wapo polls for Michigan and North Carolina.  The killer confounder region, shaded in blue, represents the sensitivity values where the estimated margin (D - R) would be opposite of the substantive finding from our estimated vote margin; the blue line represents values of $\{\rho_{\varepsilon, Y}, R^2_\varepsilon\}$ where $b^* = 0$. There are two key takeaways to highlight from the bias contour plots. First, for Michigan, we see that even if the omitted variable is relatively well balanced (i.e., $R^2_\varepsilon = 0.05$ on the $y$-axis of the Michigan plot), if the error is highly aligned with the outcome, i.e. correlated with the outcome by more than 0.5 on the $x$-axis, the omitted variable would be a killer confounder. Similarly, if the error in the weights is not highly aligned with the outcome (i.e., $\rho_{\varepsilon, Y} = 0.05$), but is highly imbalanced (i.e., $R^2_\varepsilon = 0.75$), the confounder would also be a killer confounder. This showcases the importance of considering \textit{both} sensitivity parameters when assessing the plausibility of a killer confounder.

Second, we see that the killer confounder region for North Carolina is larger than for Michigan. As such, consistent with the robustness value, there is a greater degree of sensitivity to an omitted confounder altering the estimated result in North Carolina than in Michigan. 

\subsection{Formal Benchmarking} 
Both bias contour plots and the robustness value are useful methods for summarizing overall sensitivity in a point estimate. However, in practice, it is challenging to assess whether or not they are substantively meaningful. In the context of our running example in Michigan, it can be difficult to answer whether or not it is plausible for an omitted confounder to be strong enough to explain 11\% of the variation in both the ideal weights and the outcome process. Similarly, it is challenging to know from visual inspection the plausibility of the killer confounder region. 

To address these challenges, we propose a procedure that allows researchers to use observed covariates to benchmark potential parameter values. Furthermore, we introduce several measures of relative confounding strength to help researchers assess sensitivity. 

\begin{definition}[Benchmarked Error.]\label{def:benchmarkerror}
Define a benchmark error term for observed covariate $j$ as: 
$$\hat \varepsilon^{-(j)} := w^{-(j)} - w,$$
where $w^{-(j)}$ represents the weights estimated using all covariates in $\phi(\bX)$, except for the $j$-th covariate (i.e., $\phi(\bX^{-(j)})$).
\end{definition}   

For example, for \emph{education} $\hat \varepsilon^{-(j)}$ is the error, defined as the difference between our estimated weights and those estimated omitting \emph{education}. In this setting, $\hat \varepsilon^{-(j)}$ captures the residual imbalance in educational attainment, after accounting for all other demographic variables.

We consider the confounding strength of an omitted confounder in terms of both the variation explained in the true weights and the alignment of the error and outcome. Formal benchmarking allows researchers to estimate the parameter values for an omitted confounder with equivalent confounding strength to an observed covariate, defined below. We can extend this to estimate the parameter values for an omitted confounder with confounding strength proportional, but not equivalent, to the benchmarked confounding strength. See Supplementary Materials~B.2.

\begin{definition}[Equivalent Confounding Strength.]
An omitted confounder $U$ has equivalent confounding strength to an observed covariate $\bX^{(j)}$ if: 
\begin{align}
    \var_\Sc(\varepsilon)/\var_\Sc(w^*) &= \var_\Sc(\hat \varepsilon^{-(j)})/\var_\Sc(w^*) \label{ecs_1}\\
    \cor_\Sc(\varepsilon, Y) &= \cor_\Sc(\hat \varepsilon^{-(j)}, Y) \label{ecs_2}
\end{align}
where $\hat \varepsilon^{-(j)}$ is defined in Definition~\ref{def:benchmarkerror}
\end{definition}

An omitted confounder with equivalent confounding strength explains the same amount of variation in the true weights as an observed covariate $\bX^{(-j)}$, given $\phi(\bX^{-(j)})$ (Equation~\ref{ecs_1}) and has the same level of alignment to the outcome as the benchmarked covariate (Equation~\ref{ecs_2}).

Using formal benchmarking, we can estimate the sensitivity parameters as:
\begin{equation} 
\hat R^2_\varepsilon = \frac{\hat R^{2-(j)}_\varepsilon}{1+ \hat R^{2-(j)}_\varepsilon}, \ \ \ \ \ \  \hat \rho_{\varepsilon, Y} = \hat \rho_{\varepsilon, Y}^{-(j)},
\label{eqn:benchmarking} 
\end{equation} 
where $\hat R^{2-(j)}_\varepsilon := \var_\Sc(\hat \varepsilon^{-(j)})/\var_\Sc(w)$, and $\hat \rho_{\varepsilon, Y}^{-(j)} := \cor_\Sc(\hat \varepsilon^{-(j)}, Y)$ can be directly estimated using the benchmarked error term $\hat \varepsilon^{-(j)}$. 

The benchmarked bias is estimated by plugging the values of Equation~\ref{eqn:benchmarking} to Equation~\ref{eqn:bias_decomp}.  This procedure can be extended to account for subsets of covariates, thereby allowing researchers to posit parameter values, given the aggregate confounding strength of different combinations of covariates.  Similarly, in two-step weighting approaches with a second-stage nonresponse adjustment to the weights, researchers could benchmark against the design weights.

An alternative is to estimate the \textit{minimum relative confounding strength} (MRCS), or how many times stronger (or weaker) the confounding strength must be, relative to the benchmarked covariate, in order to change the substantive direction of a point estimate.
$$\text{MRCS}(j) = \frac{\hat \mu - {b^*}}{\widehat{\text{Bias}}(\phi(\bX^{-(j)}))}.$$
An MRCS larger than 1 this implies that the omitted confounder must be stronger than the observed covariate in order to be a killer confounder. Similarly, if the MRCS is smaller than 1, then a confounder weaker than the observed covariate would be a killer confounder.

The MRCS is especially useful in cases when researchers have a strong understanding of how observed covariates relate survey sample selection and the outcome. For example, \citet{kennedy2018evaluation} found that white voters in Midwestern states with lower levels of educational attainment underrepresented in many 2016 surveys. As such, researchers conducting surveys in subsequent elections could investigate the relative confounding strength of these observed covariates. An MRCS greater than 1 implies that an unobserved confounder has to be \textit{stronger} than the confounding strength of these important observed covariates to be a killer confounder, which may be unlikely given the strength of these covariates.  In such a case we would conclude our results are insensitive. 

Benchmarking allows researchers to incorporate their substantive knowledge in understanding sensitivity. However, it is important to emphasize the limits of the benchmarked and summary measures. Our sensitivity framework allows researchers to transparently discuss the plausibility of unobserved confounders and the impact of their omission on the point estimates. However, none of our summary measures can fully eliminate the possibility of killer confounders.  In particular, while a large MRCS value may indicate a large degree of robustness, it could also mean that none of the observed covariates used to benchmark are sufficiently explanatory of the outcome or the survey inclusion probability.  This emphasizes the need for researchers to evaluate observable covariates that are highly prognostic in the diagnostics.  Without strong substantive understanding of such covariates, the sensitivity analyses here have limited diagnostic value.  Therefore, consistent with \citet{cinelli2020making}, we do not propose any thresholds or cutoff values for the robustness value or MRCS; similarly, we caution researchers from blindly using benchmarking across observed covariates, as the plausibility of fully unobserved confounders still depends on researchers' substantive judgment and context.

\subsubsection{Running Example: 2020 U.S. Presidential Election}
Figure~\ref{fig:bias_contour} contains benchmarking against observed covariates for our running example.  Each point, representing the bias of a confounder with equivalent confounding strength as the estimated benchmarked $R^2_\varepsilon$ and $\rho_{\varepsilon, Y}$, is labeled with the observed covariate.  Numeric results are presented in Table~\ref{tab:benchmarking}.

\begin{table}[ht]
\centering
\small 
\begin{tabular}{lcccc|cccccc}
  \toprule
  & \multicolumn{4}{c}{Michigan} & \multicolumn{4}{c}{North Carolina} \\
Variable & $\hat R^2_\varepsilon$ & $\hat \rho_{\varepsilon, Y}$ & MRCS & Est. Bias & $\hat R^2_\varepsilon$ & $\hat \rho_{\varepsilon, Y}$ &  MRCS & Est. Bias \\ 
  \midrule
  Party & 0.18 & -0.11 & -2.44 & -1.87 & 0.10 & -0.13 & 0.31 & -1.17 \\ 
Age & 0.26 & 0.01 & 37.63 & 0.12 & 0.18 & -0.04 & 0.74 & -0.49 \\ 
 Education & 0.34 & 0.02 & 8.75 & 0.52 & 0.36 & 0.02 & -0.71 & 0.52\\ 
  Gender & 0.04 & -0.07 & -8.14 & -0.56 & 0.12 & -0.01 & 5.07 & -0.07 \\ 
  Race & 0.13 & 0.04 & 7.86 & 0.58 & 0.04 & -0.08 & 0.81 & -0.45 \\ 
  Born Again & 0.10 & 0.04 & 10.02 & 0.46 & 0.05 & 0.13 & -0.46 & 0.80 \\ 
  \bottomrule
\end{tabular}
\caption{Formal benchmarking results for the ABC/Wapo polls for Michigan and North Carolina. The estimated bias is reported in percentage points.}
 \label{tab:benchmarking}
\end{table}

There are several key takeaways to highlight from the formal benchmarking results. First, we see that omitting a confounder with similar confounding strength as party identification, one of the strongest predictors of vote choice in U.S. politics, would result in the largest bias, given the other covariates included in weighting, understating Democratic support by 1.87 p.p in Michigan, and 1.17 p.p. in North Carolina.

Second, the benchmarking results allow us to consider the full range of the sensitivity parameters. For example, in the North Carolina poll, omitting a variable similar to born-again Christian results in a large alignment $\rho_{\varepsilon, Y}$ (0.13), but a relatively low $R^2_\varepsilon$ (0.05). This implies that while this variable is well balanced across the survey sample and the target population, the error is, as expected, highly explanatory of outcome.  Variables with similar confounding strength could result in significant bias.  In contrast, benchmarking against age results in a fairly large $R^2_\varepsilon$ ($0.18$), indicating high imbalance, but the error has a low correlation with the outcome. These covariates represent the types of confounders that, if omitted, would result in reasonably large magnitudes of bias.

Finally, the MRCS estimates have large magnitudes in Michigan,  meaning an omitted confounder would have to be substantially  (more than two times) stronger than any of the observed covariates to result in enough bias to alter the estimated outcome that Biden wins the popular vote, our chosen $b^*$.  Different choices of $b^*$ would change the MRCS. In contrast, we see that in North Carolina, the MRCS estimates for several covariates are less than 1. In particular, an omitted variable need only be 30\% as strong as party identification, or 70-80\% as strong as age or race/ethnicity to be a killer confounder. 

\paragraph{Takeaway} From the sensitivity analysis, the Michigan results are quite robust; the error from a confounder would have to explain a large degree of variation in either the outcome and the ideal weights to overturn the prediction that Biden wins the popular vote in Michigan. For example, a confounder would have to be over twice as strong as party identification, one of the strongest predictors of vote choice in American politics. In contrast, a relatively weak confounder would change the estimated results in North Carolina. 

\section{Concluding Remarks}\label{sec:conclude}
The social sciences rely heavily on surveys to answer a broad range of important questions.  In the face of rising nonresponse and growing reliance on convenience samples, survey weights are a powerful tool that allow researchers to address nonrepresentative survey samples.  Our proposed suite of sensitivity analysis tools allow researchers to reason about potential bias due to partially and fully unobserved confounders.  This includes tools for estimating bias, summary statistics and graphical analyses with formal benchmarking against observed covariates.

This paper addresses the sensitivity of a point estimate to omitting a confounder, and as such, does not explicitly account for changes in the uncertainty estimates.  Our bias decomposition holds within finite-samples, thus researchers can apply a percentile bootstrap, calculating an adjusted weighted estimate conditional on the sensitivity parameters over repeated bootstrap samples, to construct valid intervals \citep{zhao2019sensitivity, soriano2021interpretable, huang2022variance}. These confidence intervals account both for changes in the estimate from omitting a confounder as well as changes in estimated standard errors.

We focus on sensitivity to the decisions a researcher makes in the construction of survey weights.  However, there is increased emphasis on the use of outcome modeling in survey analysis, such as through model-assisted estimation \citep{breidt2017model} and doubly-robust estimation \citep{chen2020doubly}.  Our framework readily extends to such settings.  Researchers can consider the sensitivity in survey weights, given an outcome model, by replacing $Y$ in our bias decomposition with the residual of the $Y$ from the prediction from the outcome model.  See \citet{huang2021saw} for a more detailed discussion.

Finally, we note that the application in this paper is situated in a very well-studied and highly theorized substantive area. Multiple papers have evaluated the bias in public opinion polling during the 2016 and 2020 U.S. Presidential Election \citep[i.e.][to name a few]{kennedy2018evaluation, hartman2021kpop}. As a result, the example is particularly useful for discussing how to perform the sensitivity analysis. However, we note that if researchers are operating in less theorized substantive areas, reasoning about the plausibility of omitted variables may be more challenging. We emphasize that the utility of a sensitivity analysis will always be dependent on a researcher's understanding of the survey context, and cannot be used to \textit{replace} substantive knowledge. We propose a suite of tools to allow researchers to more transparently reason about sensitivity as well as better incorporate their contextual knowledge into the analysis. The sensitivity analysis and all of the corresponding sensitivity tools can be implemented using our \texttt{R} package \texttt{senseweight}.

\nocite{self_dataverse_cite}
\clearpage

\singlespacing
\printbibliography

\newpage 
\appendix 
\begin{center}
    {\Large Supplementary Materials for ``Sensitivity Analysis for Survey Weights''}
    
    Erin Hartman and Melody Huang
\end{center}
\setcounter{page}{1}
\renewcommand\thefigure{\thesection.\arabic{figure}}    
\setcounter{figure}{0}
\renewcommand\thetable{\thesection.\arabic{table}}    
\setcounter{table}{0}

\section{Glossary}

\begin{table}[h]
    \centering
    \caption{Glossary for Notation and Sensitivity Parameters}
    \begin{tabular}{c|l}
    \toprule
    \toprule
    \multicolumn{2}{l}{} \\
    \multicolumn{2}{l}{Notation} \\
    \midrule
        $S_i$ & Binary survey inclusion indicator \\
        $Y$ & Outcome of interest \\
        $\mathcal{S}$ & Units in the survey ($i: S_i = 1$) \\
        $\mu$ & Population mean of $Y$ \\
        $\bX$ & Set of observable auxiliary variables measured in both the survey and target population \\
        $\phi(\cdot)$ & A feature mapping of a given set of covariates \\
        $V$ & Partially observed variable measured in the survey but not the target population \\
        $U$ & Fully unobserved variable \\
        $w$ & Weights estimated using $\bX$ \\
        $w^*$ & Ideal weights estimated using partially or fully observed confounders \\
    
    \multicolumn{2}{l}{} \\
    \multicolumn{2}{l}{Sensitivity Parameters and Values} \\
    \midrule        
        $\varepsilon$ & The error in the estimated weights ($w - w^*$) \\
        $R^2_\varepsilon$ & Variation in ideal weights explained by the error, $\varepsilon$ \\
        $\rho_{\varepsilon, Y}$ & Correlation between the error in the weights and the outcome \\
        $\var_\Sc(w)$ & Variance of the estimated weights \\
        $\var_\Sc(Y)$ & Variance of the outcome $Y$ in the survey \\
        $b^*$ & Substantive threshold against which to evaluate robustness \\
        $RV_{b^*}$ & Robustness value at substantive meaningful $b^*$ \\
        
    \end{tabular}
    \label{tab:glossary}
\end{table}

\section{Extended Discussion}
\subsection{Calibration Weights}\label{app:calib} 
In general form, calibration weights are defined in definition~\ref{def:cal} below.

\begin{definition}[Calibration Weights] \mbox{}\\
Calibration weights $w$ are defined as:\label{def:cal}
\begin{alignat}{3}
    & \min_{w}              & \qquad& D(w, q) \label{eqn:cal_1}\\
    & \text{subject to}     &       & \sum_{i : S_i = 1} w_i f(X_i) = T, \label{eqn:cal_2}\\
    &                       &       & \sum_{i : S_i = 1} w_i = 1, \text{\ and \ } 0 \leq w_i \leq 1. \label{eqn:cal_3}
\end{alignat}
\end{definition}

\noindent where $q_i$ refers to a reference or base weight, commonly defined as unity or using survey design weights, and $D(\cdot, \cdot)$ corresponds to a distance metric which is usually greater for weights that diverge more severely from the base weight.\footnote{Formally, $D(\cdot,\cdot)$ is a divergence, not a distance, since it is often not always symmetric in its arguments.}  There are many ways to encode moment constraints, $f(X_i)$, with common methods such as ``raking'' typically using marginal population averages defined using observable characteristics, i.e. $f(X_i) = \frac{1}{N} \sum X_i$. The moment constraints defined by $f(X_i)$ in the target population are encoded in $T$.  See \citet{hartman2021kpop} for a thorough discussion of the choice of population moment constraints in calibration.

Common types of survey weighting correspond to different distance metrics $D(\cdot, \cdot)$, and are closely related to generalized regression estimation \citep{Sarndal:2007vj}.  We rely on $D(w, q) = \sum_{i: R = 1} w_i log (w_i / q_i)$ as commonly employed in ``raking'' methods and entropy balancing \citep{hainmueller2012entropy}.  
Other common weighting methods, such as post-stratification and generalized regression estimation, map to alternative distance metrics \citep{Deville:1992bv}. A full review of calibration is beyond the scope of this paper, but readers are directed to \citet{Kalton:2003vh}, \citet{Sarndal:2007vj}, \citet{Wu:2016ba}, \citet{caughey_elements_2020},or \citet{hartman2021kpop} for a more thorough treatment. The constraints in Equation~(\ref{eqn:cal_3}) jointly ensure the weights fall on the simplex; removing this constraint allows for extrapolation beyond the observed data.

An alternative approach to weighting includes inverse propensity score weighting in which weights $w_i \propto \frac{1}{\Pr(S_i = 1 \mid \bf{X})}$ \citep{little2019statistical}.  These weights make the sample representative, in expectation, on observed characteristics whereas calibration will enforce a break in the relationship between observed characteristics and $S_i$ in every sample \citep{yiu2018covassoc}.  Calibration weights converge asymptotically, with appropriate loss functions, to inverse propensity weights; for example weights estimated using logistic regression are asymptotically equivalent to raking weights using the same covariates \citep[e.g. see]{ben2021balancing}.  Raking, and most calibration estimators, are asymptotically equivalent to generalized regression estimators \citep{Deville:1992bv}.

\subsubsection{Sensitivity Analysis for Partial Confounding with Calibration Weights}

\begin{alignat}{3}
    & \min_{w}              & \qquad& D(w, q)\nonumber \\
    & \text{subject to}     &       & \sum_{i \in \mathcal{S}} w_i f(\bX_i) = T, \nonumber \\
    &                       &       & \sum_{i \in \mathcal{S}} w_i f(V_i) = \textcolor{blue}{T_V}, \label{eqn:new_constraint} \\
    &                       &       & \sum_{i \in \mathcal{S}} w_i = 1, \text{\ and \ } 0 \leq w_i \leq 1.\nonumber 
\end{alignat}
where Equation \ref{eqn:new_constraint} is the new constraint for $V_i$. By varying $T_V$, which contains the population moment constraint defined by $f(V_i)$, across a plausible range, researchers can re-estimate the survey weights over the range and re-estimate the population mean to assess how the point estimate varies as $T_V$ changes. When using raking on the margins, $T_V = \E(V_i)$. 
imbalance on a standardized scale (e.g., in terms of $z$-scores) to help provide a more intuitive understanding of what a large or small degree of imbalance is for a given covariate.

\subsection{Formal benchmarking with relative confounding strength} \label{app:benchmark} 
We now introduce formal benchmarking with relative confounding strength. To begin, define $k_\sigma$ and $k_\rho$ as follows: 
\begin{equation} 
k_\sigma = \frac{\textcolor{blue}{\var_\Sc(\varepsilon_i)/\var_\Sc(w_i^*)}}{\var_\Sc(\varepsilon_i^{-(j)})/\var_\Sc(w_i^*)}, \ \ \ \ \ \ k_\rho = \frac{\textcolor{blue}{\cor_\Sc(\varepsilon_i, Y_i)}}{\cor_\Sc(\varepsilon_i^{-(j)}, Y_i)},
\label{eqn:k} 
\end{equation}
where the blue terms are equivalent to the unobserved terms from Theorem \ref{thm:bias}. $k_\sigma$ compares the relative residual imbalance after accounting for $\bX_i$ in the unobserved confounder to the imbalance after accounting for $\bX_i^{-(j)}$ for the covariate $\bX_i^{(j)}$. Consider, for example, if educational attainment is used to benchmark the fully unobserved confounder of late decision in candidate choice. If researchers believe that late decision in candidate choice, even after adjusting for the other covariates, is more imbalanced than education, after adjusting for all covariates except for education, then $k_\sigma > 1$. However, if researchers believe that there is less residual imbalance in the confounder than the observed covariate, $k_\sigma < 1$. Setting $k_\sigma =1$ evaluates the case in which the omitted confounder has the same level of residual imbalance as the benchmarked covariate. 

Similarly, $k_\rho$ compares how correlated the outcome and the imbalance in the unobserved confounder are, relative to the correlation between the outcome and the imbalance in the benchmarked covariate. If $k_\rho > 1$, then this implies that the unobserved confounder's imbalance can explain more variation in the outcome variable than the benchmarked covariate. 

Given $k_\sigma$ and $k_\rho$, we can re-write the sensitivity parameters as functions of $k_\sigma$ and $k_\rho$ and observable quantities: 
\begin{equation} 
R^2_\varepsilon = \frac{k_\sigma \cdot R^{2-(j)}_\varepsilon}{1+ k_\sigma \cdot R^{2-(j)}_\varepsilon}, \ \ \ \ \ \  \rho_{\varepsilon, Y} = k_\rho \cdot \rho_{\varepsilon, Y}^{-(j)},
\label{eqn:k_benchmarking} 
\end{equation} 
in which $R^{2-(j)}_\varepsilon$ and $\rho_{\varepsilon, Y}$ are defined in Equation \ref{eqn:benchmarking}. 

Similar to the MRCS, researchers can solve for the minimum $k_\sigma$ and $k_\rho$ value for each observed covariate (or subset of covariates) which would result in the parameters taking on the robustness value (i.e., $\rho^2_{\varepsilon, Y} = R^2_\varepsilon = RV_q$). We denote this as $k_\sigma^{min}$ and $k_\rho^{min}$.

\section{Proofs and Derivations} 
\subsection{Corollary \ref{cor:ipw}}

To provide some intuition for $\varepsilon$, consider the following corollary. 

\begin{corollary}[Error Decomposition for Inverse Propensity-Score Weights] \label{cor:ipw}
The $\varepsilon_i$ for IPW weights estimated using $\phi(\bX_i)$ can be written as follows: 
$$\varepsilon_i = \underbrace{\frac{P(S_i = 1)}{P(S_i = 1 \mid \phi(\bX_i))} }_{\text{Estimated weights } w_i}  \underbrace{\left( 1-\frac{P(U_i \mid \phi(\bX_i))}{P(U_i \mid \phi(\bX_i), S_i = 1)} \right)}_{\text{Residual imbalance in } U_i}$$
More specifically, $\varepsilon_i$ is a function of the estimated weights (left), and the residual imbalance in the omitted confounder, after accounting for $\phi(\bX_i)$ (right). 
\end{corollary} 

When using inverse propensity weights, the estimated weights and the ideal weights are : 
$$w_i = \frac{P(S_i = 1)}{P(S_i = 1 \mid \phi(\bX_i))} \ \ \  \ \ \ \ w^*_i = \frac{P(S_i = 1)}{P(S_i = 1 \mid \phi(\bX_i), U_i)}$$
Then the error term is: 
\begin{align*} 
\varepsilon_i &= w_i - w_i^* \\
&= \frac{P(S_i = 1)}{P(S_i = 1 \mid \phi(\bX_i))} - \frac{P(S_i = 1)}{P(S_i = 1 \mid \phi(\bX_i), U_i)} \\
&= P(S_i = 1) \cdot \left( \frac{1}{P(S_i = 1 \mid \phi(\bX_i))} - \frac{1}{P(S_i = 1 \mid \phi(\bX_i), U_i)} \right)
\intertext{Applying Bayes' Rule:}
&=  P(S_i = 1) \cdot  \left(\frac{1}{P(S_i = 1 | \phi(\bX_i))} - \frac{P(U_i | \phi(\bX_i))}{P(U_i | \phi(\bX_i),S_i = 1) \cdot P(S_i = 1 | \phi(\bX_i))} \right)\\
&= \frac{ P(S_i = 1) }{P(S_i = 1 | \phi(\bX_i))} \cdot \left( 1 - \frac{P(U_i \mid \phi(\bX_i)}{P(U_i \mid  \phi(\bX_i),S_i = 1)} \right) 
\end{align*} 
In cases when the omitted confounder $U_i$ is binary, we can rewrite the expression as a function of the conditional average of $U_i$: 
$$\varepsilon_i = \frac{ P(S_i = 1) }{P(S_i = 1 | \phi(\bX_i))} \cdot \left( 1 - \frac{\E(U_i \mid \phi(\bX_i)}{\E(U_i \mid  \phi(\bX_i),S_i = 1)} \right) $$
\subsection{Theorem \ref{thm:bias}}
\begin{align} 
\text{Bias}(\hat \mu) &= \E(\hat \mu) - \mu \nonumber \\
&= \E \left( \sum_{i \in \Sc} w_i Y_i \right) - \mu \nonumber \\
&= \E \left( \sum_{i \in \Sc} w_i Y_i \right) - \E \left( \sum_{i \in \Sc} w_i^* Y_i \right) & \text{by conditional ignorability assumption}\nonumber \\
&= \E_\Sc(w_i Y_i) - \E_\Sc(w_i^* Y_i) & \text{by linearity in expectation} \nonumber \\
&= \E_\Sc((w_i- w_i^*) \cdot Y_i)\nonumber \\
&= \E_\Sc((w_i- w_i^*) \cdot Y_i) - \E_\Sc(w_i - w_i^*) \E_\Sc(Y_i) &\text{by construction, $\E(w_i) = \E(w_i^*)$}\nonumber \\
&= \E_\Sc(\varepsilon_i \cdot Y_i) - \E_\Sc(\varepsilon_i) \cdot \E_\Sc(Y_i) \nonumber \\
&= \cov_\Sc(\varepsilon_i, Y_i) \nonumber \\
&= \cor_\Sc(\varepsilon_i, Y_i) \cdot \sqrt{\var_\Sc(Y_i) \cdot \var_\Sc(\varepsilon_i)}
\label{eqn:bias_decomp1} 
\end{align} 
Defining $R^2 = \frac{\var_\Sc(\varepsilon_i)}{\var_\Sc(w_i^*)}$ and noting that $\var_\Sc(w_i^*)$ can be written as $\var_\Sc(w_i) + \var_\Sc(\varepsilon_i)$: 
\begin{align} 
\var_\Sc(\varepsilon_i) &= \var_\Sc(w_i^*) - \var_\Sc(w_i) \nonumber \\
&= \var_\Sc(w_i^*) \underbrace{\left( 1 - \frac{\var_\Sc(w_i)}{\var_\Sc(w_i^*)} \right)}_{\equiv R^2_\varepsilon} \nonumber 
\intertext{Then, noting that $\var_\Sc(w_i^*) = \var_\Sc(w_i) /(1-R^2_\varepsilon)$}
&= \var_\Sc(w_i)  \cdot \frac{R^2_\varepsilon}{1-R^2_\varepsilon} 
\label{eqn:var_eps} 
\end{align} 
Substituting Equation \eqref{eqn:var_eps} into Equation \eqref{eqn:bias_decomp1} recovers the results from Theorem \ref{thm:bias}. 

\textbf{Remark.} It is worth noting that the bias formula and subsequent sensitivity analyses derived are implicitly treating the estimated weights $w$ and the ideal weights $w^*$ as fixed. This is mathematically equivalent to looking at the asymptotic bias of the weighted estimators (see \citet{huang2021saw}). However, we can extend the same framework and bias expressions for the finite-sample case, in which we define the bias as the error in the estimated weights $\hat w$ and a set of oracle weights $\hat w^*$ (i.e., see \citet{cinelli2020making} and \citet{soriano2021interpretable} as examples). We provide extensions for relaxing this assumption in Section \ref{sec:conclude}.

\subsection{Corollary \ref{cor:var_decomp}}
The results of Corollary \ref{cor:var_decomp} follow results from \citet{huang2021saw}, who show that for inverse propensity weights, projecting the ideal weights $w_i^*$ into the space of observed covariates $\phi(\bX_i)$ recovers the estimated weights $w_i$  (i.e., $\E(w_i^* \mid \phi(\bX_i) ) = w_i$). As such: 

\begin{align*} 
\var_\Sc(\varepsilon_i) &= \var_\Sc(w_i - w_i^*) \\
&= \var_\Sc(w_i) + \var_\Sc(w_i^*) - 2 \cov_\Sc(w_i, w_i^*) \\
&= \var_\Sc(w_i) + \var_\Sc(w_i^*) - 2 \left( \E_\Sc(w_i \cdot w_i^*) - \E_\Sc(w_i) \E_\Sc(w_i^*) \right)
\intertext{Because $\E_\Sc(w_i) = \E_\Sc(w_i^*) = 1$ and by Law of Iterated Expectation:}
&= \var_\Sc(w_i) + \var_\Sc(w_i^*) - 2 \left( \E_\Sc(\E_\Sc(w_i \cdot w_i^* | \phi(\bX_i))) - \E_\Sc(w_i)^2 \right) 
\intertext{Because $\E_\Sc(w_i^* | \phi(\bX_i) ) = w_i$:}
&= \var_\Sc(w_i) + \var_\Sc(w_i^*) - 2 \left( \E_\Sc(w_i^2) - \E_\Sc(w_i)^2 \right) \\
&= \var_\Sc(w_i) + \var_\Sc(w_i^*) - 2 \var_\Sc(w_i) \\
&= \var_\Sc(w_i^*) - \var_\Sc(w_i)
\end{align*} 
As such, several immediate properties follow. First, by definition of variance, $\var_\Sc(\varepsilon_i) = \var_\Sc(w_i^*) - \var_\Sc(w_i) \geq 0$, which implies that $\var_\Sc(w_i^*) \geq \var_\Sc(w_i)$. Second, defining $R^2_\varepsilon := \var_\Sc(\varepsilon_i)/\var_\Sc(w_i^*)$, we see that $R^2_\varepsilon$ is naturally bound on the interval [0,1]. 
\section{Partially Observed Confounders}

\subsection{Sensitivity Analysis for Partially Observed Confounders with Inverse Propensity Weighting\label{app:ipw_partial}}

We now consider an example where $V$ is binary and researchers estimate inverse propensity score weights.  In this case, we can rewrite the results of Corollary~\ref{cor:ipw} as follows. 

\begin{example}[IPW Weights with Binary $V$.]\label{ex:ipw_partial}
When $V$ is binary, we can rewrite the error term, $\epsilon$, for IPW weights from Corollary~\ref{cor:ipw} as a function of sample and population proportions of $V$ conditional on $\phi(\bX)$ as
$$\varepsilon_i = \underbrace{\frac{P(S_i = 1)}{P(S_i = 1 \mid \phi(\bX)_i)}}_{:= w_i} \cdot \left( 1 - \frac{\textcolor{blue}{\E(V_i \mid \phi(\bX)_i)}}{ \E(V_i \mid S_i = 1, \phi(\bX)_i) } \right).$$
\end{example}

The key takeaway from Example~\ref{ex:ipw_partial} is that, because $V$ is observed across the sample, the denominator $\E(V_i \mid S_i = 1, \phi(\bX)_i)$ can be directly estimated from the observed data, reducing the problem to a single sensitivity parameter.  The sensitivity analysis requires researchers to posit reasonable values of the unobservable $\E(V_i \mid \phi(\bX)_i)$, denoted in blue, estimate $\varepsilon$, and adjust the point estimate via Theorem~\ref{thm:bias}.  This is similar to the sensitivity analysis for partial confounders proposed by \citet{nguyen2017sensitivity}, which uses an outcome model based approach.

Calibration only requires that researchers reason about the unconditional population quantity, since it will implicitly solve for the weights that meet the balancing constraints without needing to know the conditional expectations.  However, in some simple cases it may be possible for researchers to reason about the conditional means, for example if the estimated weights only include a handful of covariates with a limited number of levels.  In general, though, this is a difficult quantity to reason about without rich auxiliary data, and if the researcher had such rich auxiliary data, they would likely not have a partially observed confounder.

\subsection{Detection of Partially Observed Confounders \label{app:est_partial_confound}}

While researchers can often posit partially observed confounders based on theoretical considerations, sometimes researchers are unsure if there are observed covariates that may be partially observed confounders.  In this section we propose a method for detecting partially observed confounders that can be used for determining the existence of a partially observed confounder or for confirming if a theoretically relevant covariate is a partially observed confounder.  These variables can then be used in the one parameter sensitivity analysis described in Section~\ref{sec:partial_observe}.

We consider the data setting in which researchers can posit a sampling set--all variables related to sample selection--either through data driven methods or from theory, but this set is only partially observed, i.e. every variable is measured in the survey sample and all but one variable is measured in the target population. Throughout this section, we will denote $\bX^S$ as the sampling set, and $\bX$ as the entire set of observed variables. We assume, without loss of generality, that there is only one partially observed covariate, $V$ and return to this assumption below.\footnote{Formally, this means $\bX^S \subseteq \{\bX \cup V\}$ and $\{\bX \cap V\} = \varnothing$, with $\bX^S$ measured for all units with $S_i = 1$.  When all covariates in $\bX^S$ are fully observed, i.e. $\bX^S \subseteq \bX$, our proposal reduces to a variable selection method for constructing survey weights.}  Because $V$ is not measured in the target population, we cannot rely on an appropriate feature expansion $\phi(\bX^S)$ of the full sampling set to construct survey weights. 

We adapt an estimation technique from \citet{egami2019covariate}. The method, described in detail in Appendix~\ref{app:est_partial_confound}, uses the survey data to estimate a Markov Random Field \citep{yang2014mixed} (MRF), or an undirected graph, over the variables in $\{\bX^S, \bX \not\subset \bX^S, Y\}$. We then determine if there is a set of variables that render the outcome and all variables in the sampling set $\bX^S$ conditionally ignorable, thus justifying Assumption~\ref{assump:cond_ig}, but which does not include the partially observed confounder.  If no such set exists, then $V$ is a partially observed confounder.  In this case, we suggest researchers use the sensitivity analysis for partially observed confounders described in Section~\ref{sec:partial_observe}.  When there are multiple partially observed confounders, the algorithm can also be updated to determine the set that contains the smallest number of partial confounders as an additional constraint.

To begin, we first formalize the idea that if a separating set consisting of fully observed covariates exist, then this set may be used in lieu of the partially observed selection set to identify the population mean. 

\begin{theorem}[Identification of Population Mean from Survey Sample \citep{egami2019covariate}]\label{thm:marginal} 
When $\bX^S$ is known and partially observed in the survey sample, for units $S_i = 1$, consider a set $\bW$ that is fully observed, then under Assumption~\ref{assump:cond_ig} (replacing $\bX$ with $\bW$):
\begin{equation}
    Y_i \ \indep \ \bX^S_i \mid \bW_i, S_i = 1 \implies Y_i \ \indep \ S_i \mid \bW_i
\end{equation}
\end{theorem}

\textbf{Proof:}
\begin{align}
        & Y_i \indep \bX^S_i \mid \bW_i, S_i = 1 & \\
        \implies & Y_i \indep \bX^S_i \mid \bW_i, S & \quad\text{by ignorability} \label{thm:marg:2} \\
        & \text{Note: } Y_i \indep S_i \mid \bW_i, \bX^S_i & \quad\text{By definition of Sampling Set} \label{thm:marg:3}\\
        \implies & Y_i \indep {S_i, \bX^S_i} \mid \bW_i & \quad\text{Combining Equations~(\ref{thm:marg:2}) and (\ref{thm:marg:3})} \\
        & & \quad\text{using Intersection principle in \citet{pearl2000}} \nonumber \\
        \implies & Y_i \indep S_i \mid \bW_i & \qed
\end{align}

\noindent The intuition behind Theorem~\ref{thm:marginal} is that if a separating set $\bW$ renders all variables in the sampling set conditionally ignorable to the outcome within the survey sample, Assumption~\ref{assump:cond_ig} holds and $\bW$ can be used to identify and estimate the population mean. Because conditional ignorability need only hold across the survey sample, we can directly estimate using just the survey sample data whether or not there exists a $\bW$ that blocks all paths between the outcome $Y$ and the sampling set $\bX^{S}$. When $\bW$ does not exist, this implies that the partially observed $V$ is necessary for unbiased identification of the population mean. 

To estimate such separating sets $\bW$, we use the algorithm described in \citet{egami2019covariate}, outlined in Table~\ref{tab:summary}.  We first estimate a Markov Random Field (MRF) over all variables in $\bX^S$ (some of which are partially observed), all additional covariates that are fully observed $\bX$, and the outcome $Y$.  MRFs are statistical models that encode conditional independence structures of variables using graph separation rules. We estimate the MRFs using mixed graphical models \citep{yang2014mixed, haslbeck2020mgm}, which allow for a mixture of continuous and categorical covariates.  Figure~\ref{fig:application_MRF} contains an estimated MRF for our application to the 2020 U.S. Presidential survey conducted by ABC News/Washington Post for Michigan.

\begin{table}[!h]
\small
  \begin{center}
    \begin{tabular}{r|p{5.25in}}
      \toprule
      \multicolumn{2}{l}{\textbf{Estimating Separating Sets \citep{egami2019covariate}} } \\
      \midrule
      Step 1: & Using mixed graphical models, estimate a Markov Random Field over $\{\bX^{S} \cup \bX\}$ and the outcome $Y$. Define $q$ as the total number of covariates. Store all simple paths from $Y$ to $\bX^S$ from the estimated MRF in a matrix $\bP$.\\ \vspace{-5mm}  \\[8pt]
      Step 2: & Define $\bf{v}$ to be a $q$-dimensional vector that encodes which variables are partially observed, where ${\bf{v}}_j = 1$ if the $j$th variable is only measured in the survey sample and not the target population, and 0 otherwise. \\ \vspace{-5mm} \\[8pt] 
      Step 3: & Solve the following linear programming problem: 
      \begin{equation}
        \min_{\bf{d}} \sum_{j = 1}^{q} d_j \quad \text{s.t.,\ } \bP\bf{d} \ge \bf{1} \text{ and } \bf{v}^\top\bf{d} = 0
        \label{eqn:optim}
        \end{equation}
        \textit{If a fully observed, valid separating set $\bW$ is found:}\\
         &$\qquad$\rotatebox[origin=c]{180}{$\mathbf{\Lsh}$} Use construct survey weights using estimated sparating set $\bW$.\\
        &\textit{If no fully observed separating set exists:}\\
        &$\qquad$\rotatebox[origin=c]{180}{$\mathbf{\Lsh}$} Partial confounders exist.  Estimate weights with observed covariates and conduct sensitivity analysis for partially observed confounders (Section \ref{sec:partial_observe}).
      \\
      \bottomrule
    \end{tabular}
    \end{center}
    \caption{Summary of the algorithm for detecting partially observed confounders. We refer readers to \citet{egami2019covariate} for more details on how to estimate the MRF.}\label{tab:summary}
\end{table}

Once we estimate the MRF, we can solve for the separating set as a constrained, linear programming problem, in which we optimize for a separating set of smallest size.\footnote{Alternative loss functions include the set that yields the least dispersion in weights or the lowest variance in the point estimate.} We constrain the optimization problem such that (1) that all paths between $Y$ and the variables in $\bX^S$ are blocked by the candidate separating set (i.e., $\bP \textbf{d} \geq 1$, in Equation \ref{eqn:optim}), and (2) that the candidate separating set does not contain  the partially observed variable $V$ (i.e., $\bf{v}^\top \bf{d} = 0$, in Equation \ref{eqn:optim}), where $\bf{v}$ encodes which variables are observed in the target population. If the optimization problem returns a valid separating set, then this implies that partially observed confounding is not a concern. However, if no feasible set exists, then this implies that given the fully observed variables available to the researcher for constructing survey weights, there is no set of weights that can recover the population mean without bias due to $V$.

\subsubsection{Running Example: 2020 U.S. Presidential Election} \label{subsubsec:example_mrf} 
We return now to our running example.  We assume our fully observed covariates include $\bX =$\,\{\textit{Age}, \textit{Gender}, \textit{Race/Ethnicity}, \textit{Education Level}, \textit{Party ID}, \textit{Born-Again Christian}\}.  As in Section~\ref{subsec:running_example}, these demographic variables are commonly used to construct survey weights.  Existing literature indicates that political interest is also correlated with propensity to respond to surveys; however it is not commonly incorporated into survey weights because it is not available in target populations data.  Therefore, we let $V =$\{\textit{\mbox{Political Interest}}\}, encoded as whether individuals are "very closely" following the upcoming 2020 U.S. Presidential Election.  We assume that the sampling set includes this partially observed variable as well as these common demographic variables, i.e. $\bX^S =$\,\{\textit{Age}, \textit{Gender}, \textit{Race/Ethnicity}, \textit{Education}, \textit{Party ID}, \textit{Born-Again Christian}, \fbox{\textit{Political Interest}}\}, where the box denotes that $V$ is partially observed.  The algorithm seeks to determine if there is a set of variables in $\bX$ that renders political interest and the outcome conditionally independent, thus justifying Assumption~\ref{assump:cond_ig}.

\begin{figure}
    \centering
    \includegraphics[width=0.6\textwidth]{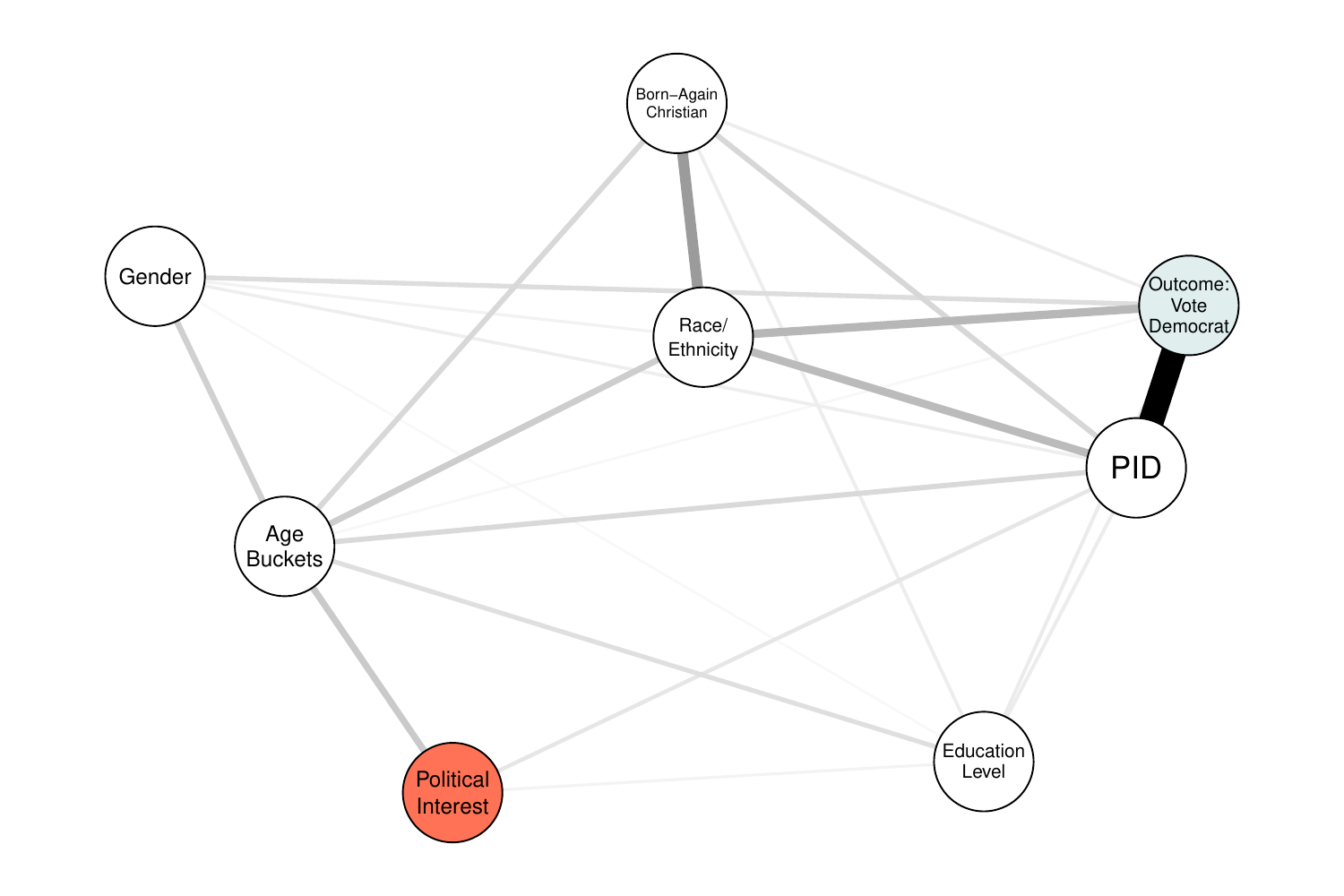}
    \caption{MRF of ABC News/Washington Post poll conducted in late October 2020 in Michigan.  White nodes indicate fully observed covariates, red nodes indicate partially observed covariates, and the gray node is the outcome.  Lines between nodes indicate a conditional correlation exists, and the width of the line indicates the strength of the correlation.}
    \label{fig:application_MRF}
\end{figure}

Figure~\ref{fig:application_MRF} presents the undirected graph estimated for the ABC News/Washington Post poll.  The algorithm returns that political interest is not a partially observed confounder, as it can be rendered conditionally independent using our weighting variables.  This is visually evident by the fact that the outcome is not connected to political interest.  If researchers choose to use only fully observed covariates in the construction of the survey weights, they need not worry about confounding from this political interest variable. 

\section{Extended Results for the Empirical Application}\label{app:nopid}

In our main analysis, we use party identification as a weighting variable, as it is the strongest predictor of vote choice in American politics and it is measured in our target population defined by the CES.  However, whether or not to weight on party identification can be controversial.  In this section, we conduct our analysis for Michigan without using party identification in our weights.  We then use party identification as a partially observed confounder to assess sensitivity.  Unsurprisingly, we see that the results are quite sensitive to the exclusion of such a strong predictor of the outcome.  This is confirmed in the bias contour plots, as well.

\subsection{Partially Observed Confounding}
To begin, we treat party identification as a partially observed variable.  In this section, we assess sensitivity to identification as a Democrat and Republican separately, for the purpose of visual exposition.  One could also weight on the full three-way category, and present the results as a heat map, which would result in similar findings.

\begin{figure}[ht]
    \centering
    \includegraphics[width=0.75\textwidth]{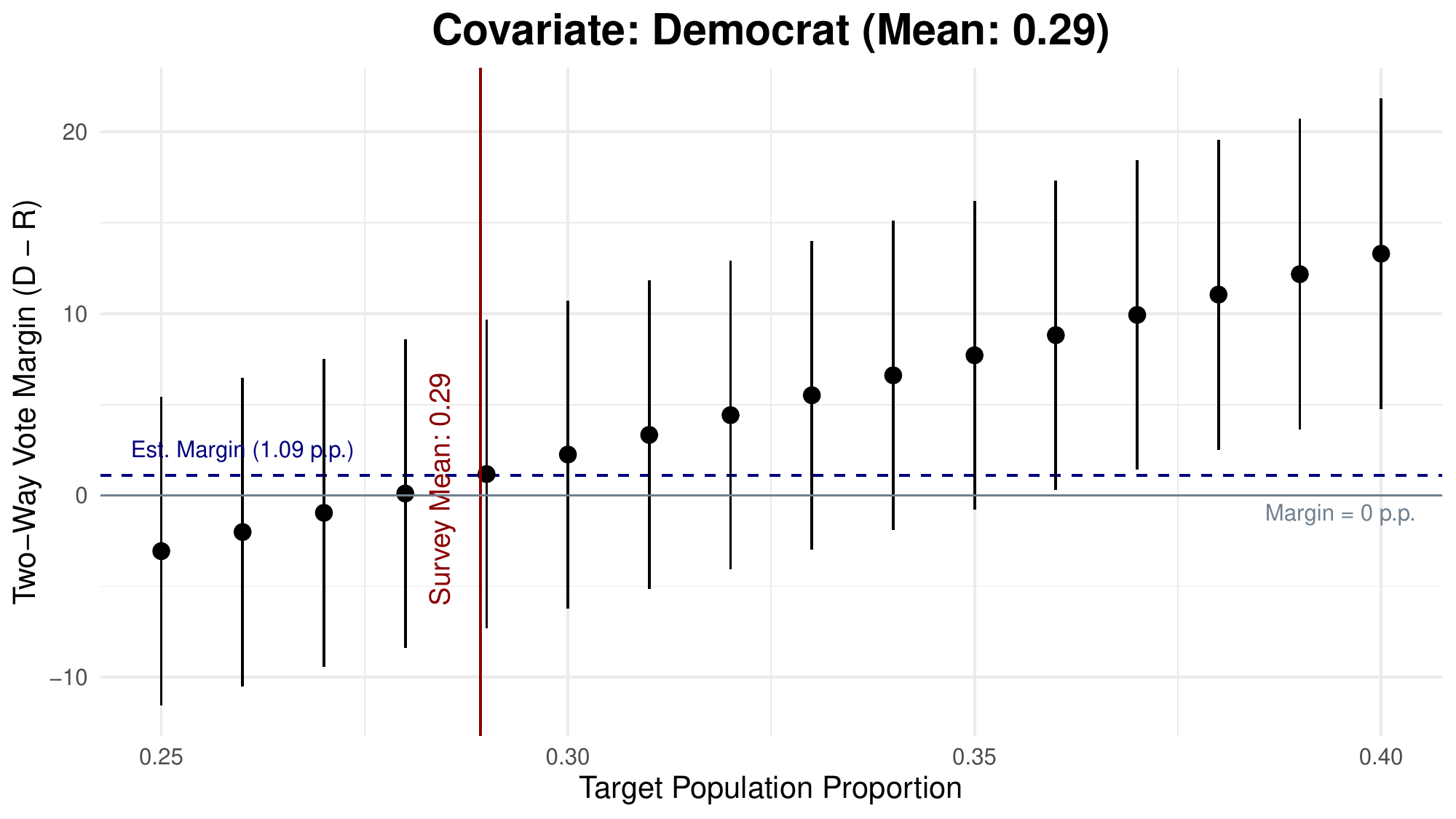}
    \includegraphics[width=0.75\textwidth]{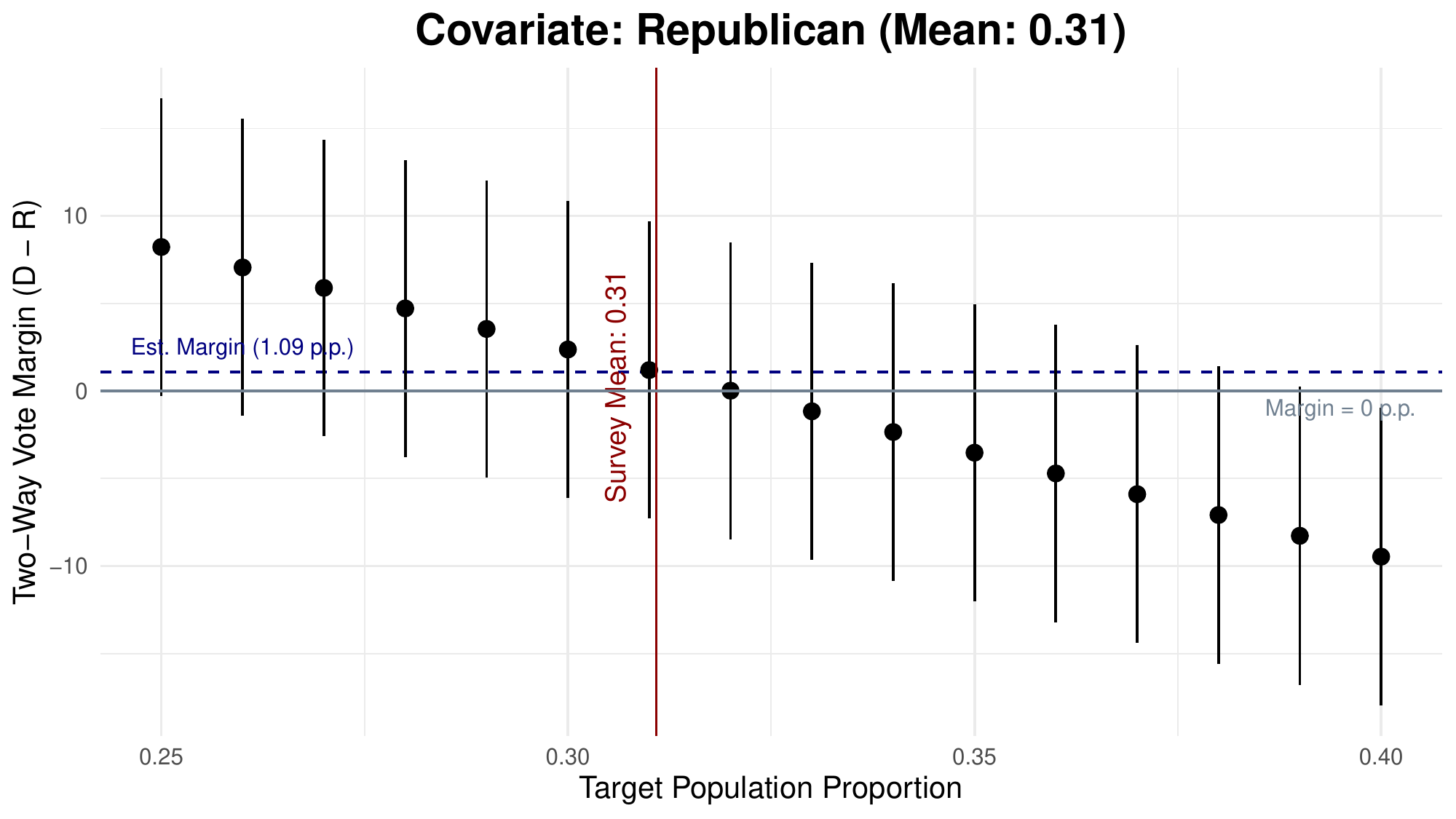}
    \caption{Partial confounding plots. We vary the target population of the partially observed covariates (i.e., individuals who identify as Democrats or Republicans) and plot the resulting estimates.}
    \label{fig:MI_no_pid_partial}
\end{figure}

We begin by assessing sensitivity to the proportion of Democrats. The weighted survey average of proportion of Democrats is 0.29 using our survey weights that exclude party identification. We vary the target population proportion of Democrats and re-estimate the vote margin of a Biden victory in Michigan. Consistent with what we expect, we see that if the population proportion of Democrats is larger than 0.29, then this means that by omitting Democrat from our weights, we have underestimated the vote margin of a Biden victory. In contrast, if the population proportion of Democrats is less than 0.29, then by omitting Democrat from the weights, we will have overestimated the vote margin of a Biden victory. We repeat this analysis for the proportion of Republicans, and find consistent results. We visualize the results in Figure \ref{fig:MI_no_pid_partial}.  We have truncated the x-axis to a "reasonable" range for these sensitivity parameters, assuming that no less than 25\% or more than 40\% of the population identifies for each party.  It is worth noting that in our target population, the true proportion of Democrats is 35\% and Republicans is 31\%.  If a researcher can reason that the proportion of Democrats is most likely understated, then these results would indicate the poll most likely understates the true margin.

\subsection{Fully Unobserved Confounding} 
We now conduct our sensitivity analysis for fully unobserved confounding. We begin by computing the robustness value, at a threshold of $b^* = 0$ (see Table \ref{tbl:MI_no_pid_summary}). We see that in contrast to the robustness value reported in the main text, by not accounting for party identification, the robustness value drops from 0.11 to 0.03. In other words, a confounder that results in an error that explains 3\% of the variation in the ideal weights and the outcome will be sufficiently strong to reduce our estimated vote margin to zero. This is consistent with what we expect, as we know that party identification controls for a lot, and by not accounting for it, our estimate may be more sensitive to potential confounders.
\begin{table}[ht]
\centering
\begin{tabular}{lccc}
   \toprule
& Weighted Estimate & SE & $RV_{b^* = 0}$ \\ 
   \midrule
Two-Way Vote Margin (D-R) & 1.09 & 3.71 & 0.03 \\ 
    \bottomrule
 \end{tabular}
 \caption{Sensitivity summary for Michigan, without accounting for party identification.}
 \label{tbl:MI_no_pid_summary} 
\end{table}

To assess the plausibility of such a killer confounder, we turn to benchmarking (see Table \ref{tbl:MI_no_pid_benchmarking} for results). We see that by omitting party identification from our analysis, omitting variables with equivalent confounding strength to gender or whether or not an individual is a born-again Christian would result in enough confounding to overturn our estimated result of a Biden victory in Michigan. Similarly, we see that omitting a confounder like age or race would also result in a large amount of bias.  Note that, without weighting on party identification, the correlation of the errors from omitting these variables is stronger than the original analysis, where the error from omitting these variables was less strong because party identification explained much of the variation with the outcome.

\begin{table}[ht]
\centering
\begin{tabular}{lcccc}
  \toprule
Variable & $\hat R^2_\varepsilon$ & $\hat \rho_{\varepsilon, Y}$& MRCS & Est. Bias \\ 
  \midrule
Age & 0.29 & 0.04 & 1.43 & 0.76 \\ 
Education & 0.31 & 0.04 & 1.10 & 1.00  \\ 
  Gender & 0.07 & -0.13 & -0.92 & -1.19 \\ 
  Race & 0.12 & 0.07 & 1.21 & 0.90 \\ 
  Born Again & 0.11 & 0.15 & 0.63 & 1.74 \\ 
   \bottomrule
\end{tabular}
\caption{Benchmarking results for Michigan, without accounting for party identification.}
\label{tbl:MI_no_pid_benchmarking} 
\end{table}

\begin{figure}[ht] 
\centering 
\includegraphics[width=0.5\textwidth]{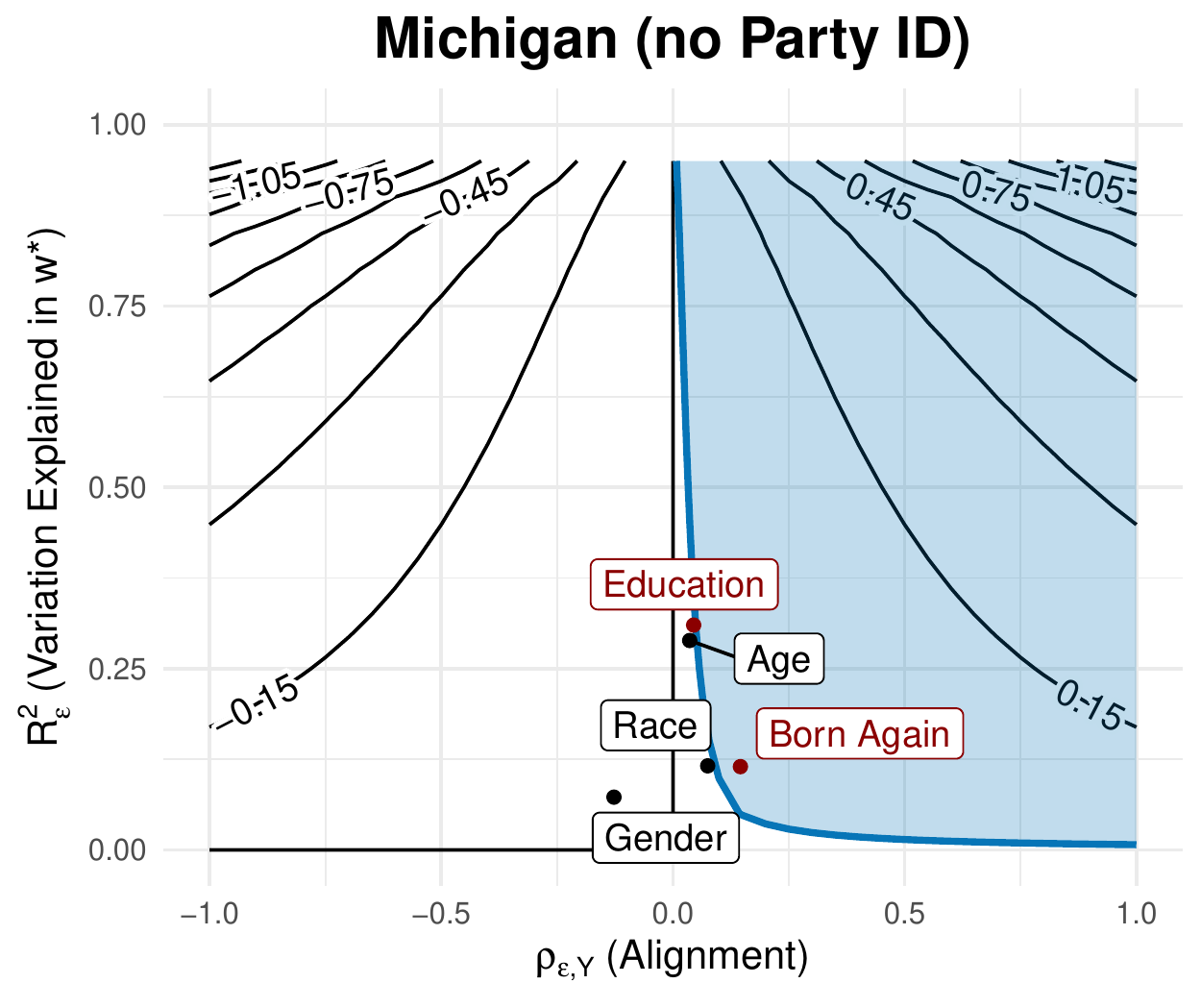}
\caption{Bias contour plot for Michigan, not accounting for party identification.}
\label{fig:MI_no_pid_contour}
\end{figure}

We also generate the bias contour plot (see Figure \ref{fig:MI_no_pid_contour}).

\end{document}